\documentclass{ws-jai}
\usepackage[flushleft]{threeparttable}
\usepackage[usenames,dvipsnames]{color}
\usepackage{graphicx}
\usepackage{afterpage}
\newcommand{\e}{$e^{-}$}
\newcommand{\eps}{$e^{-} \,$s$^{-1}$}

\begin{document}

\catchline{}{}{}{}{} 

\markboth{M.~Zemcov \textit{et al.}}{Optimal Photocurrent Estimation for Resource-Constrained Imaging}

\title{An Algorithm for Real-Time Optimal Photocurrent
  Estimation including Transient Detection for
  Resource-Constrained Imaging Applications}

\author{Michael Zemcov$^{1,2,\dagger}$, Brendan Crill$^2$, Matthew
  Ryan$^2$, Zak Staniszewski$^2$}

\address{
$^1$Center for Detectors, School of Physics and Astronomy, Rochester
Institute of Technology, Rochester, NY 14623, USA, zemcov@cfd.rit.edu\\
$^2$Jet Propulsion Laboratory, Pasadena, CA 91109, USA \\
}

\maketitle

\corres{$^\dagger$Corresponding author.}

\begin{history}
\received{2016 April 5};
\revised{2016 May 6};
\accepted{2016 May 18}
\end{history}

\begin{abstract}
  Mega-pixel charge-integrating detectors are common in near-IR
  imaging applications.  Optimal signal-to-noise ratio estimates of
  the photocurrents, which are particularly important in the
  low-signal regime, are produced by fitting linear models to
  sequential reads of the charge on the detector.  Algorithms that
  solve this problem have a long history, but can be computationally
  intensive.  Furthermore, the cosmic ray background is appreciable
  for these detectors in Earth orbit, particularly above the Earth's
  magnetic poles and the South Atlantic Anomaly, and on-board
  reduction routines must be capable of flagging affected pixels.  In
  this paper we present an algorithm that generates optimal
  photocurrent estimates and flags random transient charge generation
  from cosmic rays, and is specifically designed to fit on a computationally
  restricted platform.  We take as a case study the Spectro-Photometer
  for the History of the Universe, Epoch of Reionization, and Ices
  Explorer (SPHEREx), a NASA Small Explorer astrophysics experiment
  concept, and show that the algorithm can easily fit in the
  resource-constrained environment of such a restricted platform.
  Detailed simulations of the input astrophysical signals and detector
  array performance are used to characterize the fitting routines in
  the presence of complex noise properties and charge transients.  We
  use both Hubble Space Telescope Wide Field Camera-3 and Wide-field
  Infrared Survey Explorer to develop an empirical understanding of
  the susceptibility of near-IR detectors in low earth orbit and build
  a model for realistic cosmic ray energy spectra and rates.  We show
  that our algorithm generates an unbiased estimate of the true
  photocurrent that is identical to that from a standard line fitting
  package, and characterize the rate, energy, and timing of both
  detected and undetected transient events.  This algorithm has
  significant potential for imaging with charge-integrating detectors
  in astrophysics, earth science, and remote sensing applications.
\end{abstract}


\section{Introduction}

Modern infrared detectors have multi-mega pixel formats
\citep{Beletic2008}, so can produce large data volumes in short times
(\textit{e.g.} \citealt{Smith2012}).  This is problematic in space
applications, where data telemetry is expensive.  This problem is
typically solved by using some form of Correlated Double Sampling
(CDS; \citealt{Fowler1990}, \textit{e.g.}  \citealt{Robberto2007}),
which permits the information from multiple reads of a detector during
a single integration to be summarized by a single total pixel
number$N_{\rm pix}$-sized frame.  In contrast, ``Sample Up the Ramp''
(SUR) methods offer larger signal-to-noise ratios in a given
integration time \citep{Garnett1993}, as well as the ability to
identify and reject cosmic ray and other transient events, at the cost
of significantly increased data volumes \citep{Rauscher2007}.  In
order to implement SUR algorithms in telemetry-constrained
applications, some amount of data processing and reduction must be
performed on-board.

As a specific application, we have studied these algorithms for use in
the Spectro-Photometer for the History of the Universe, Epoch of
Reionization, and Ices Explorer
(SPHEREx\footnote{\url{http://spherex.caltech.edu}};
\citealt{Dore2014}).  SPHEREx is a proposed Small Explorer mission
designed to generate $\lambda/\Delta \lambda=41.5$ spectra for
0.75$<\lambda<$4.1$\mu$m and $R=150$ spectra for
4.1$<\lambda<$4.8$\mu$m for every 6.2 arcsecond pixel over the entire
sky.  It will utilize four H2RG detectors read out using SIDECAR ASICs
\citep{Loose2006} in 32-channel mode driven at a $\sim 1.5 \,$s sample
rate.  To cover the entire spectral band, we will use two
$2.5 \, \mu$m and two $5 \, \mu$m optimized detectors operating at
$\gtrsim 50 \,$K.  Similar systems are planned to be deployed on
upcoming large missions like the James Webb Space Telescope (JWST;
\citealt{Rauscher2007}) and Euclid \citep{Crouzet2012}, allowing us to
leverage existing characterization studies to predict the performance
of SPHEREx.  Because of the narrow spectral window passed to the
detectors, we expect median photocurrents $\lesssim 1 \,$\eps\ per
pixel.  As a result, it is crucial to optimize the signal-to-noise
ratio per integration, leading the SPHEREx team to baseline the SUR
algorithm, which has the added benefit of providing robustness against
detector transients.

The streamlined and efficient nature of Small
Explorer\footnote{\url{http://explorers.gsfc.nasa.gov/smex.html}}
class missions necessitates the use of high-heritage, resource-constrained
processing platforms, restricting the possible scope of on-board data processing
algorithms.  For similar reasons, the data telemetry
bandwidth is constrained, which places a premium on data reduction
compression.  These competing requirements impose a significant
challenge: we need to maximally reduce the transmitted data volume,
while allocating the reduction processing minimal resources.  This
paper presents our solution to this problem, which takes the form of a
set of resource-constrained algorithms that: (\textit{i}) return the
optimal SUR photocurrent estimate; (\textit{ii}) allow real-time
transient flagging; and (\textit{iii}) minimize the required long term
data storage and telemetry bandwidth.  The paper is organized as
follows: in Sec.~\ref{S:datadescription} we summarize the expected
characteristics of the SPHEREx data as they relate to the on-board
processing.  We present the data reduction algorithms in Sec.~
\ref{S:dataprocessing}, and then show a variety of characterization
and validation tests in Sec.~\ref{S:testing}.  In Sec.~
\ref{S:hardware}, we present details of the hardware implementation of
the SPHEREx flight system, and show that the algorithms fit within the
available resource envelope.  Finally, we discuss broader implications
of this work in Sec.~\ref{S:discussion}.

\section{Data Description}
\label{S:datadescription}

The SPHEREx H2RG detector arrays are \textit{charge-integrating},
which is to say, incident photons generate a current in the detector
material that is proportional to the photon flux.  However, it is the
charge on the detector pixels that is amplified and digitized by the
readout system.  The photon signal is therefore proportional to the
derivative of the measured charge on the detector over time, which
leads to the SUR estimate for the signal intensity.  The current at
the detector $i_{\mathrm{phot}} $ can be related to the intensity of
the astrophysical signal $\lambda I_{\lambda}$ via:
\begin{equation}
\label{eq:Iph}
i_{\mathrm{phot}} \simeq \lambda I_{\lambda} \left( \frac{\eta A \Omega}{h \nu}
\frac{\Delta \lambda}{\lambda} \right),
\end{equation}
where $A \Omega$ is the pixel throughput, $\eta$ is the total
efficiency, and $\Delta \lambda$ is the integral bandwidth
(\textit{e.g.} \citealt{Bock2013}).  As a result, SPHEREx operates in
a point-and-stare observational mode.  The dominant astrophysical
signal in an average array pixel is due to Zodiacal Light (ZL) which
has a known brightness versus wavelength \citep{Kelsall1998}, allowing
us to tailor our integration lengths to be photon-noise limited.  In
this study, we baseline the all-sky survey integration time
$T_{\rm int}$ per field to be $\sim 107\,$s per pointing so that the
photocurrent noise is always dominated by shot noise from photons.
Given this $T_{\rm int}$, the spectral resolution of the instrument,
and the orbit, it is possible to calculate an observation sequence
that will tile the entire sky in six months \citep{Spangelo2015}.
  
\begin{figure*}[ht]
\centering
\includegraphics*[width=6.5in]{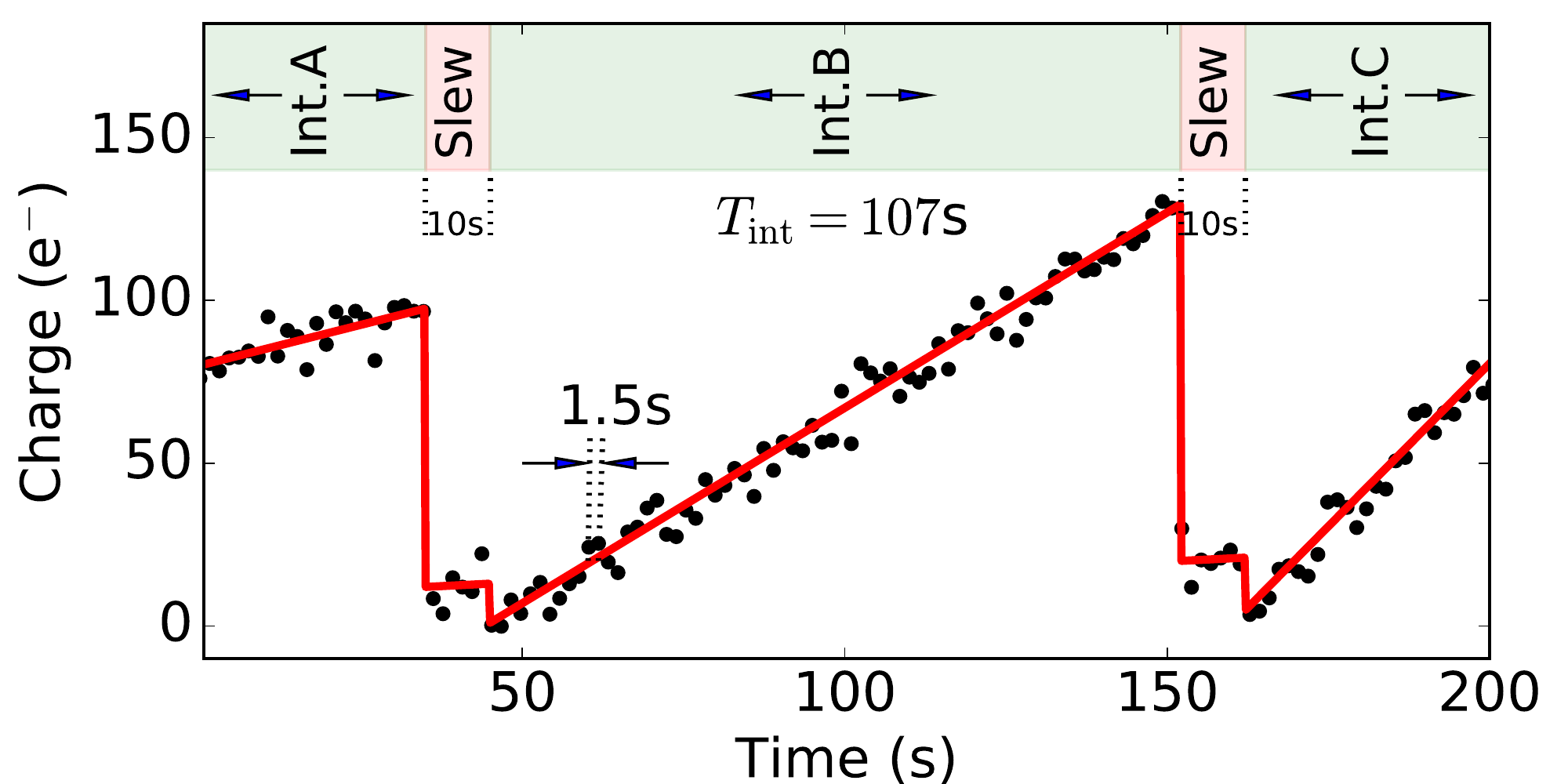}
\caption{Schematic view of a typical SPHEREx observing sequence.  The
  plot shows the end of an observation (`Int.~A'), a slew to a new
  target, a full $107 \,$s integration (`Int.~B'), a slew to a third
  target, and the beginning of a third integration (`Int. C').  Data
  samples are indicated with a $1.5 \,$s sample rate (black points),
  and the input photocurrent in each section is also shown (red line).
  The steps at the beginning and end of each integration are due to
  detector resets. \label{fig:qvst} }
\end{figure*}

A given observation consists of a slew to the target field, an array
charge reset, integration on the target for $T_{\rm int}$, followed by
an array reset and then a slew to the next target.  During the
integration, charge values are sequentially read from the array with a
$\Delta t = 1.5 \,$s pixel visit cadence and accumulated into various
sums (see Sec.~\ref{S:dataprocessing}).  At the end of the
integration, the algorithm solves for the photocurrent, compresses the
output, and stores it to disk.  An schematic time series is shown in
Fig.~\ref{fig:qvst}, highlighting the sequence.

H2RG detectors exhibit a rich phenomenology of noise that has already
been well studied.  \citet{Rauscher2015} has built code for a noise
model which captures much of the salient phenomenology (see
Sec.~\ref{sS:input} for details).  In this work, we assume the SPHEREx
arrays have the parameters shown in Table \ref{tab:arrays}.  Missing
from this model and required to complete our tests is a model for
cosmic ray events, which can have a large effect above the shielding
of the earth's atmosphere.

\begin{table}[ht]
\centering
\caption{Assumed H2RG array parameters. \label{tab:arrays}}
\begin{tabular}{lc}
\hline
Parameter & Value \\ \hline

Array type & Teledyne H2RG \\

Array format & $2048 \times 2048$ pixels \\ 

Output mode & 32 channel \\

Frame read rate & $1.5 \,$s \\

Dark current $i_{\rm d}$ & 5 m\eps/pixel \\

Read noise & 5.2 \e \\

DC pedestal drift rms & 4 \e \\

Correlated pink noise rms & 3 \e \\

Uncorrelated pink noise rms & 1 \e \\

Correlated alternating column noise rms & 0.5 \e \\ 

Equivalent CDS rms & 10.5 \e \\ \hline

\end{tabular}
\end{table}

\subsection{Cosmic Ray Rate and Spectrum}
\label{sS:cosmicrays}

There is little data available on the \textit{in situ} cosmic ray
susceptibility of modern, large-format HgCdTe arrays deployed in
space.  No instrument currently in flight uses H2RG-class detectors;
however, both the Hubble Space Telescope (HST) Wide-Field Camera-3
(WFC-3; \citealt{Dressel2016}) and the Wide-field Infrared Survey
Explorer (WISE; \citealt{Wright2010}) instruments use H1RG detectors,
which are essentially a single quadrant of an H2RG detector.  In
addition, the mean altitudes of both HST ($560 \,$km) and WISE
($525 \,$km) are similar but slightly lower than that predicted for
SPHEREx ($600 \,$km), putting them in comparable regions of the
earth's radiation belts.  Finally, both the WFC-3 and SPHEREx
detectors are substrate-removed, ensuring the cosmic ray
cross-sections are similar.

Of course, there are complications to this comparison, not least of
which is the difference in cosmic ray shielding between WFC-3, WISE,
and SPHEREx.  However, without a detailed physical model of the
instruments that models the material properties, geometries,
\textit{etc.}, it is not possible to account for these differences.
For the purpose of this study, we choose to neglect these differences.
As shown below, the cosmic ray rates are low enough that even an order
of magnitude error from this assumption would typically have a
negligible effect on the SPHEREx survey.

To complete this study, we have used a set of dark data both in
``regular'' regions of the earth's magnetic field, and while in the
South Atlantic Anomaly (SAA), a region in the magnetosphere where
cosmic ray event rates are known to be dramatically higher.  The data
sets are summarized in Table \ref{tab:data}.  For this study, we use
offset-subtracted array reads separated by a given time per read,
calibrated to \eps\ using known gain factors.

\begin{table}
\small
\centering
\caption{Data sets used in the WFC-3 cosmic ray study.  \label{tab:data}}
\begin{tabular}{lccccc}
\hline
Field Type & Acquisition Dates & Number of files & Reads per file & Integration time
per read ($dt$) & Total exposure time \\ \hline

Regular & 2009 - 2016 & 140 & 14 & 200 s & 2800 s \\ 

SAA & Jun.~-Jul.~2009 & 84 & 3 & 50 s & 150 s$^{\ast}$ \\ \hline

\multicolumn{6}{l}{$^{\ast}$ The integration time in the SAA region is
  necessarily shorter as HST passes through the region on a similar
  time scale.} \\ \hline

\end{tabular}
\end{table}

The data analysis proceeds assuming a simple model for cosmic ray
events.  Pixels (which in this case have no external illumination)
have a small linear ramp with time corresponding to the dark current,
which is $\sim$m\eps\ for this detector.  The read noise is much
larger than this, on the order of electrons, so a
typical pixel's time stream is effectively just noise with
approximately no gradient.  This means the array currents should be
well modeled by a histogram with mean very close to zero, and shape
symmetric about that mean\footnote{Note it is not necessary to assume
  that the noise properties are particularly well-behaved: we only
  require noise pathologies occur in both the positive-charge and
  negative-charge directions equally.}.  To isolate the photocurrent,
we subtract the first frame read from the last, which removes
instrumental offsets.  The result, when corrected for the integration
time, is the charge accumulated on a detector over the exposure time.
The effect of cosmic rays is to skew the positive-going side of a
histogram of the array pixel values to larger values.

Pathological pixels are problematic for this analysis as they can
deviate strongly from the behavior of a normal pixel, and could look
like either (\textit{a}) strongly negative-going events, which would
skew our estimate of the non-white component of the noise, or
(\textit{b}) strongly positive-going events, appear to be cosmic
rays.  In fact, the number of pathological pixels could be much larger
than the typical number of cosmic ray events, so these must be
eliminated.  To develop a bad pixel mask, we perform an analysis where
we compute differences between randomly selected exposure pairs.  Bad
pixels are those defined as \emph{always} displaying a pathology, so
should be the same between exposures separated by a random amount of
time.  We perform this procedure with a selection of 10 random
pairings without replacement, and require that bad pixel candidates
are flagged in a minimum of two different data set pairs to be flagged
as bad in the final mask.  From this selection, we form a pixel mask
that should have identified most of the pathological pixels in the
detector.  In all, $1.7 \,$\% of pixels are flagged and ignored in the
following analysis.

\begin{figure*}[ht!]
\centering
\includegraphics*[width=3.2in]{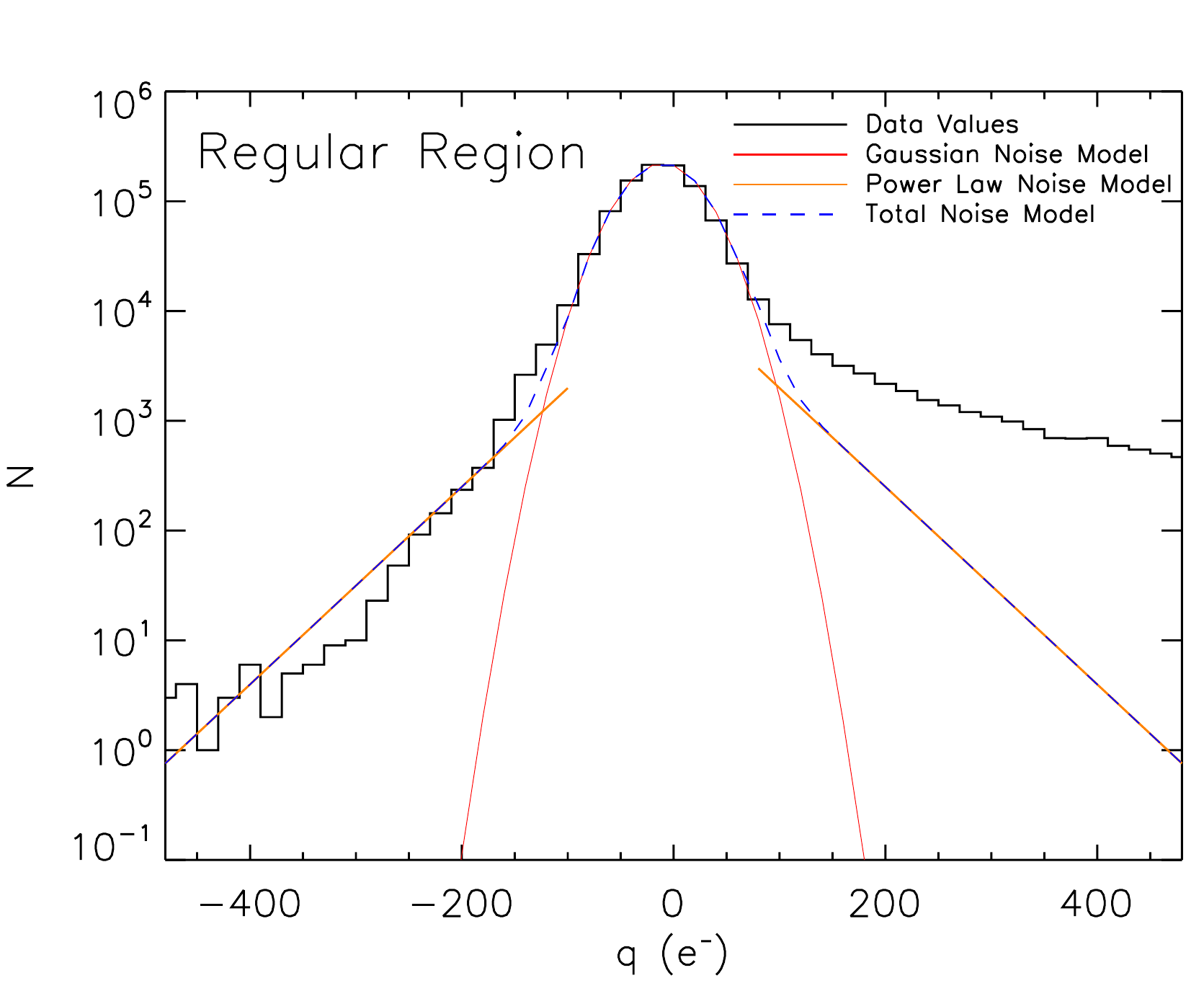}
\includegraphics*[width=3.2in]{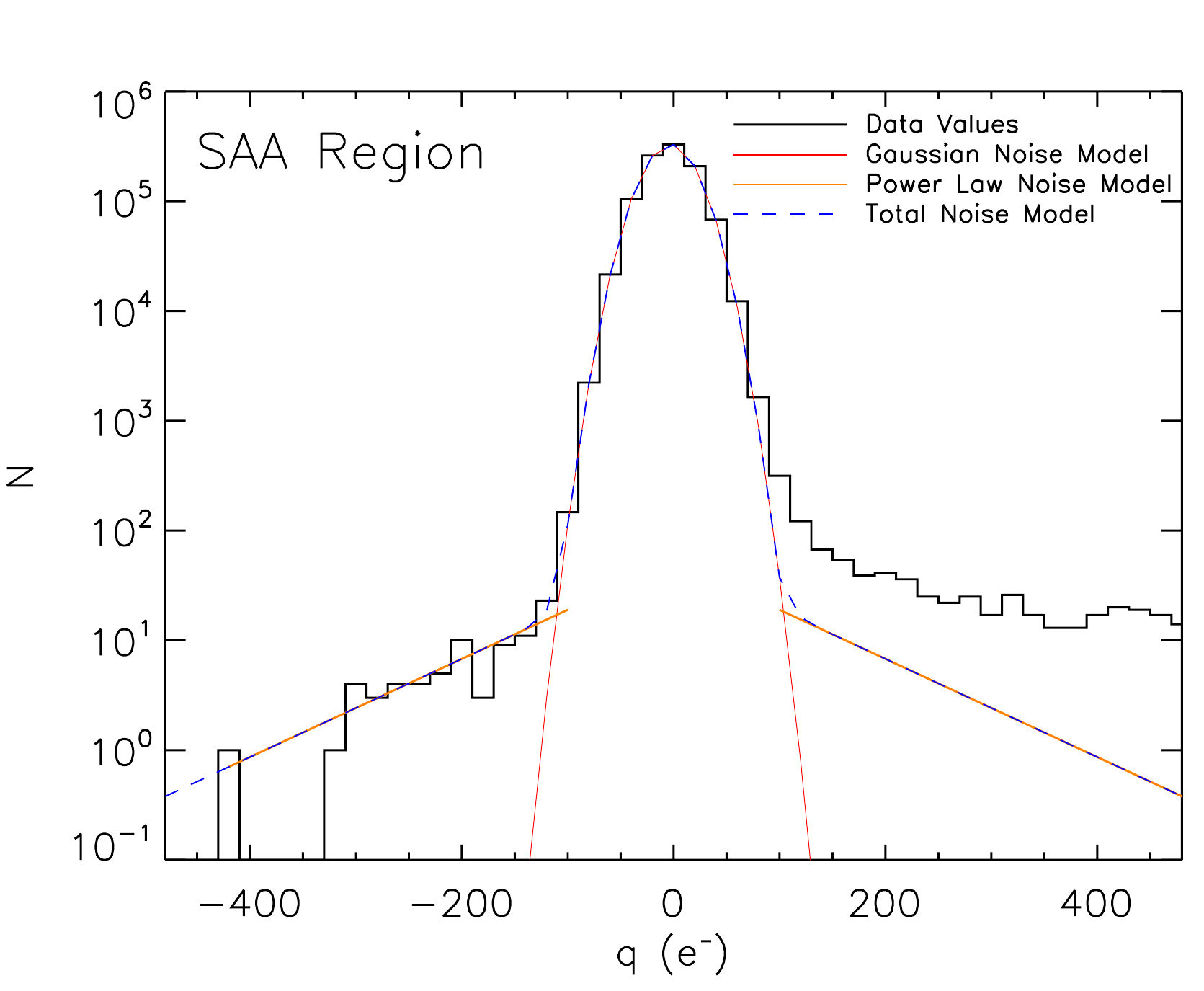}
\caption{Histograms of the pixel values for two files typical of the
  SAA and regular regions. The two data sets have been scaled to the
  common SAA integration time of $150 \,$s.  This has the effect of
  causing the white-noise component width to be the same between the
  two sets, but increases the relative contribution of rare events in
  the non-SAA data.  Also shown are a Gaussian model for the quasi-white
  noise contribution (solid red line) with an rms noise of
  $\sim 24 \,$e$^{-}$ (corresponding to $\sim 150 \,$me$^{-}$/s).  To
  account for the non-Gaussian population of pixels creating the
  negative-going wing, we fit a power law model (yellow line), and
  also show the sum of the noise components (blue line).  Assuming
  that the non-Gaussian noise component is equal in both the negative-
  and positive-going directions, the excess of the histogram on the
  positive-going side over the power law term is due to cosmic ray
  events. \label{fig:histograms} }
\end{figure*}

Having flagged bad pixels, we can compute histograms of the remaining
pixels to find cosmic ray hits, shown in Fig.~\ref{fig:histograms},
including models for the Gaussian and power law-like noise terms.  As
the detector noise should be symmetric about the mean, the power law
derived from the negative-going data should also match the
positive-going side.  In both the non-SAA and SAA cases, an excess of
positive-going events is evident at $q \gtrsim 100 \,$e$^{-1}$, which
we associate with cosmic ray events.

\begin{figure*}[ht!]
\centering
\includegraphics*[width=3.2in]{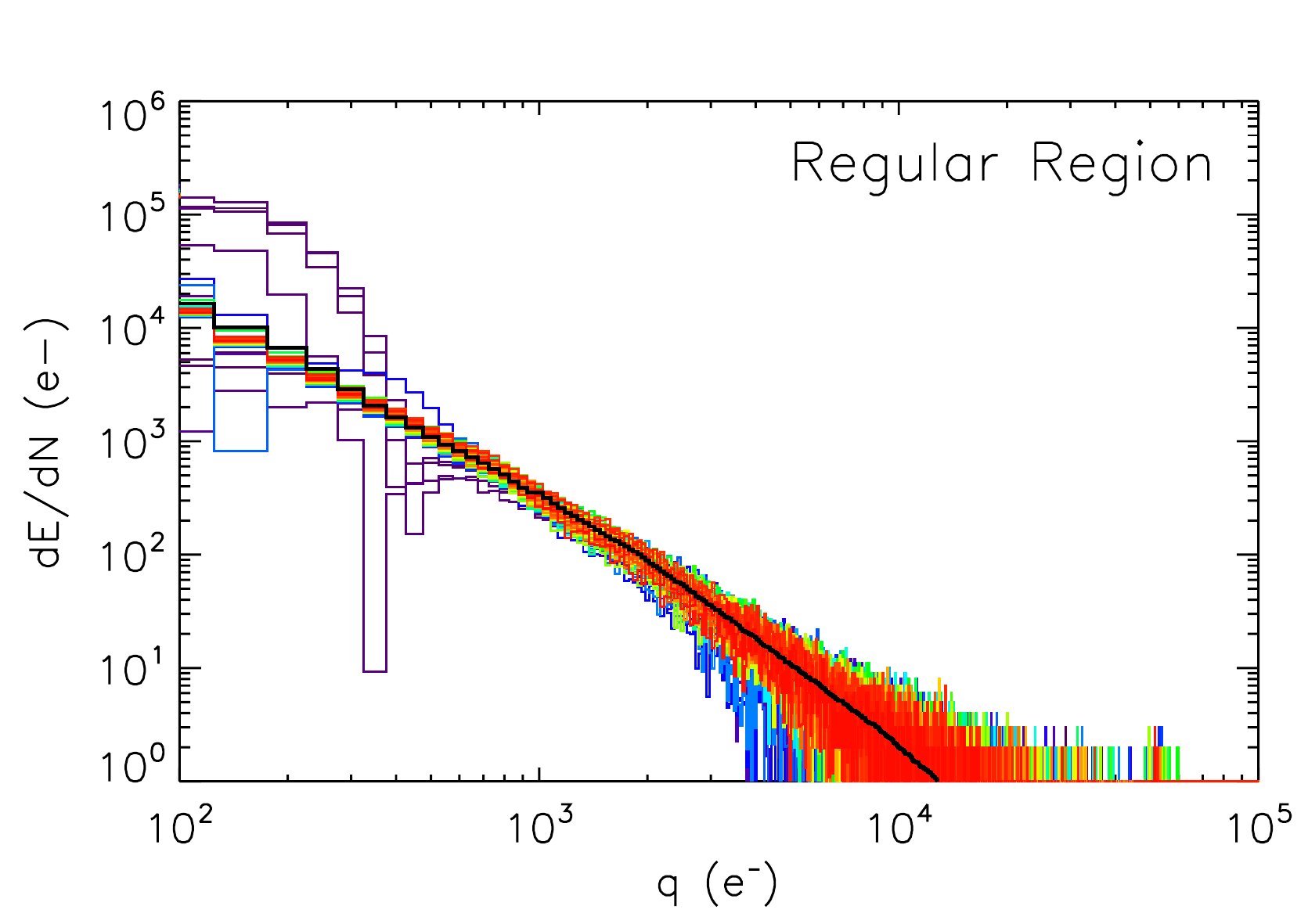}
\includegraphics*[width=3.2in]{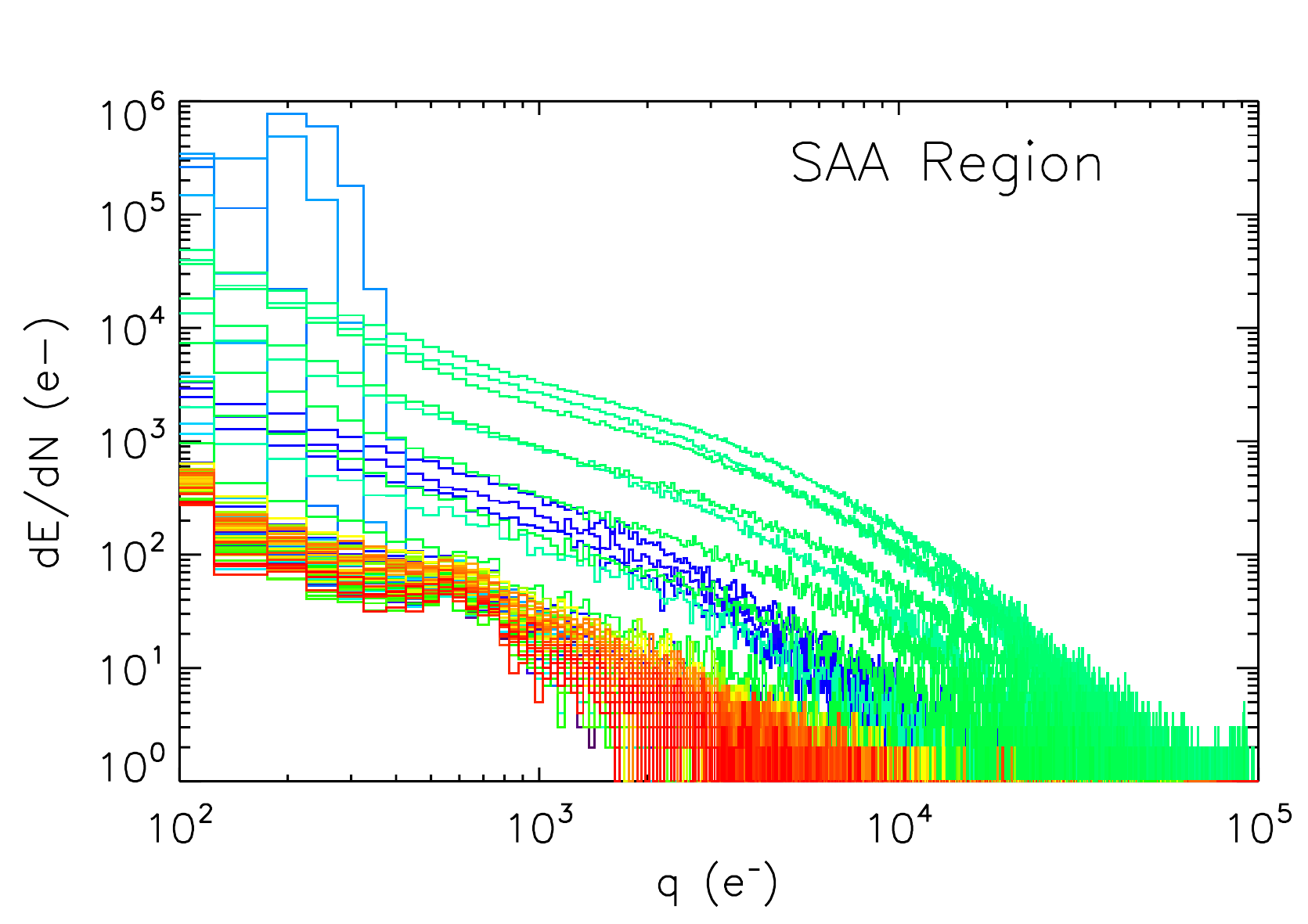}
\caption{Cosmic ray energy spectra for both regular and 
  (left panel) and SAA regions (right panel).  Colors correspond to
  data set number, given in Table \ref{tab:data}.  \label{fig:spectrum} }
\end{figure*}

To isolate these cosmic ray events, we subtract the noise model curve
from the histogram.  In Fig.~\ref{fig:spectrum} we show the
resulting energy spectrum for cosmic rays as measured in the two data
sets.  The regular region data are remarkably consistent, and follow a
smooth power law model.  The SAA region data are less consistent, but
the lowest-event rate sets are very similar to the regular region
data.  There are, however, large outlier events that are not rare.  It
is not possible from these data alone to determine how these outlier
data sets correlated with external factors such as path through the
SAA or solar activity, though it is probably a reasonable assumption
that such correlation is present.  From these data, we determine an
event rate that is $1.3 \times 10^{-4}$ events s$^{-1}$ pixel$^{-1}$,
consistent with previous analyses of similar data \citep{Barker2010}.

Though the WFC-3 detector is a good analog to those of SPHEREx, HST
does not follow a polar orbit, and as a result does not pass through
the earth's magnetic poles.  These may trap cosmic rays similarly to
the SAA region, and potentially could have an even larger event rate
or more energetic spectrum.  To place limits on this, we use cosmic
ray hit data from WISE, which follows a very similar earth-terminator
orbit to SPHEREx.  These data give cosmic ray event rates over the
full sky at a $535 \,$km altitude near beginning of the WISE
mission.  The WISE event rates are qualitatively consistent with those
of WFC-3 in the SAA.  Scaling to the SPHEREx integration time, we
would expect $\sim 1 \%$ percent of pixels to suffer from detectable
events when within 30 degrees of latitude of each pole, corresponding
to $1/6^{\rm th}$ of an orbit, is a modest increase in the total
number of events over the baseline expectation.

\section{Data Processing Algorithm}
\label{S:dataprocessing}

The algorithm described here is developed from the ``normal equation''
formalism\footnote{So named because the equations
  specify that the residual must be normal to every vector in the span
  of $A$.} (see \textit{e.g.} \citealt{Press1992} chapter 15 for a review), and
shares common heritage with algorithms developed to address similar
requirements in previous NASA flight projects (\citealt{Fixsen2000},
\citealt{Offenberg2005}).  Because of the limited resources available
to SPHEREx, we have developed this algorithm to optimize against the
following metrics:
\begin{enumerate}
\item \textit{Return the optimal\footnote{Here we use ``optimal'' in
      the sense of maximum signal-to-noise ratio in a given
      integration time, rather than providing an optimal estimate of
      the \emph{error} in the fit parameters.} best-fitting
    photocurrent}, while
\item \textit{robustly detecting cosmic rays and other transients}, and
  simultaneously 
\item \textit{minimizing RAM requirements}, and
\item \textit{minimizing the number of computations}.
\end{enumerate}
We place no emphasis on returning accurate error estimates for the
fits, or for retrieving the full-integration slope in cosmic ray-affected
pixels using a constrained fit model\footnote{This is in constrast to
  \textit{e.g.} \citet{Offenberg2005} where the on-board processing
  algorithm is designed with these considerations in mind.}.

\subsection{Processing Architecture}
\label{sS:dataarchitecture}

\begin{figure*}[ht!]
\centering
\includegraphics*[width=6.5in]{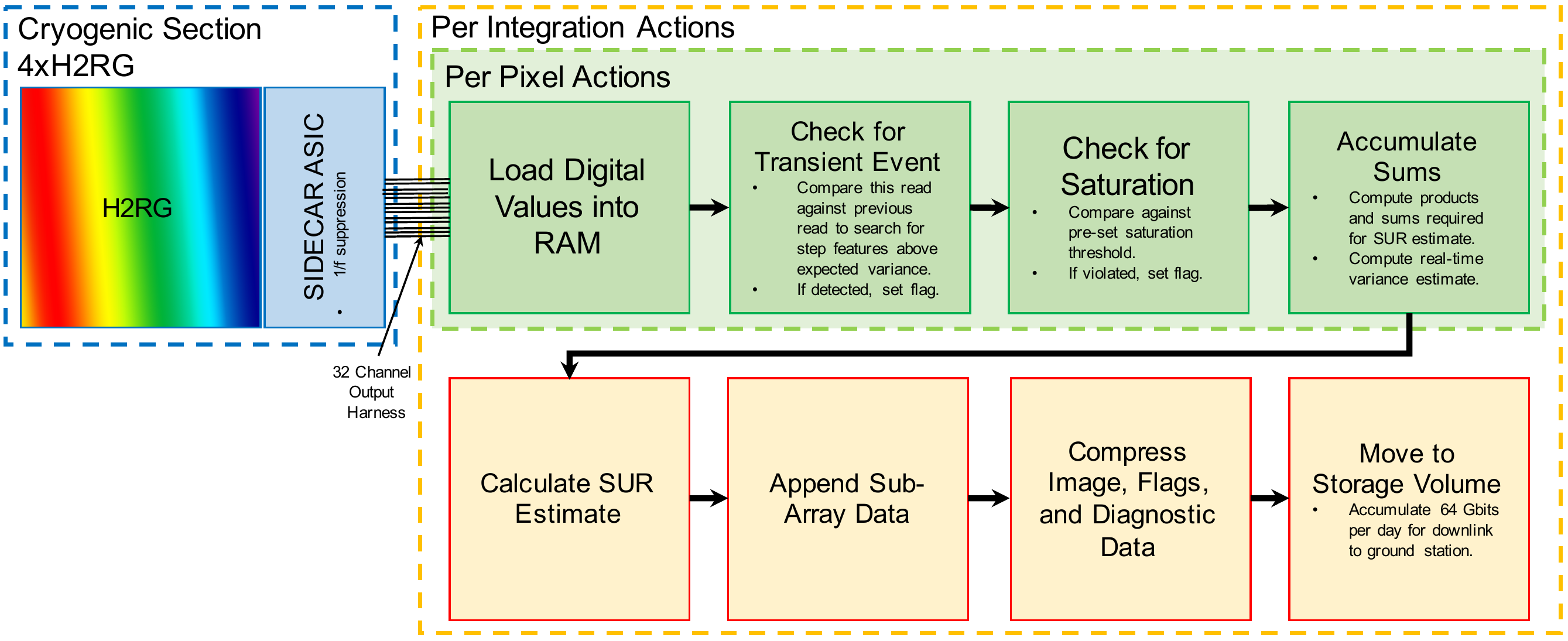}
\caption{On-board data processing flow for SPHEREx.  For each
  detector, data is sampled from the H2RG using 32 output channels and
  loaded into RAM on the processing board.  At each time step, we
  check and flag for time stream transient events and pixel
  reciprocity failure before accumulating various quantities (see
  Sec.~\ref{sS:SUR}).  At the end of each integration, these
  quantities are then used to build optimal estimates for the
  photocurrent in each detector.  A $8 \times 32$ pixel sub-array
  sampled at full time resolution is appended to the image and flag
  data, which are then compressed.  These are then shifted to a
  storage volume for later transmission to the ground
  station. \label{fig:architecture} }
\end{figure*}

The overall architecture of the on-board processing pipeline for
SPHEREx is summarized in Fig.~\ref{fig:architecture}.  At each time
sample for each pixel, the data is loaded into the on-board RAM.
Next, the algorithm checks for transient events and detector
saturation using the algorithms detailed in the next section.
Finally, various quantities are computed and updated before the next
sample arrives.  Following the accumulation of a full integration, the
SUR estimate is computed using the summed information and pre-computed
look-up values.  For SPHEREx, we are concerned about the performance
of the on-board processing algorithms over the lifetime of the
mission, so also append the full time series for an $8 \times 32$
sub-image in each of the 4 H2RG detector arrays.  This sub-array
information can be analyzed in detail during later analysis and the
results compared against the in-flight processing results to monitor
the algorithm performance.  This is also useful for determining the
cosmic ray hit rate and energy spectrum in the as-built instrument, as
well as checking the pixel performance stability over the lifetime of
the mission.  Finally, the data set comprising the SUR image, flags,
and diagnostic data are compressed and moved to storage for later
telemetry to the ground station.

\subsection{SUR Photocurrent Estimates}
\label{sS:SUR}

In the most general form, a linear relationship between the
photocurrent $\partial q/\partial t$ and reads of the charge on the
detector $q$ can be written as:
\begin{equation}
\label{eq:qi}
q_{i} = A_{ij} \cdot m_{j} + n_{i}
\end{equation}
where $i$ corresponds to sample number, $j$ to pixel
  number, $m_{0} = f \propto \partial q/\partial t$ is the incident
flux in suitable units, $m_{1} = q_{0}$ the initial charge on the
detector, $A_{ij}$ the $i \times 2$ design
matrix\footnote{In statistics, a ``design matrix'' is
    a matrix of explanatory variables of a set of objects. In this
    particular instance, each row represents an individual time
    sample, with the successive columns corresponding to pixel samples
    at that time.} encoding the linear transform, and $n_{i}$ is a
random number drawn from a suitable probability density function (PDF)
to match the statistics of the physical process at hand (here the sum
of photon noise and read noise, which can potentially be arbitrarily
complex functions of $f$ and $q$).  In a fitting problem, we want to
solve for the value of $m$, which is given by:
\begin{equation}
\label{eq:linefit}
  m = (A^{\rm T} N^{-1} A)^{-1} A^{\rm T} N^{-1} q,
\end{equation}
where $A$ is the design matrix, $^{\rm T}$ is the transpose operation, and
$N$ is the noise covariance matrix computed from $n$.  

To simplify the most general solution, we can take advantage of the
fact that SPHEREx will stare at an observation field with a fixed
sampling cadence and pre-determined $T_{\rm int}$.  Because most
pixels will have photocurrents $\sim 1 \,$\eps, the noise term will be
close to constant in a given integration (see Sec.~
\ref{ssS:photocurrent}).  This allows us to set
$\sigma_{i}^{2} = \sigma_{j}^{2}$ so the matrix is diagonal.  This
means that the $N$ terms appear as factors of $\sigma^{-2}$ in the
$(A^{\rm T} N^{-1} A)^{-1}$ and factors of $\sigma^{2}$ in the
$A^{\rm T} N^{-1}$ term, and cancel.  Thus $N$ drops from the problem.

The fixed sampling cadence imposes a well-defined, regular structure on the
design matrix, while the pre-defined integration time means the design
matrix's ultimate size is known.  The structure of the design matrix
is:
\begin{equation}
  A = \begin{bmatrix} 0 & 1 \\
    1 & 1 \\
    2 & 1 \\
    \vdots & \vdots \\
    n-1 & 1
  \end{bmatrix} = [i,1] \\
\end{equation}
where $n$ is the number of points in the integration.  As a result,
the term $(A^{\rm T} A)$ is simply a $(2 \times 2)$ matrix composed
as:
\begin{equation}
 (A^{\rm T} A) = \begin{bmatrix} \sum_{0}^{n-1} A_{0i}^{2} &
    \sum_{0}^{n-1} A_{0i} \\
    \sum_{0}^{n-1} A_{0i} & \sum_{0}^{n-1} 1 \end{bmatrix}
\end{equation}
To solve Equation \ref{eq:linefit}, we need to compute
$(A^{\rm T} A)^{-1}$ from this.  Each computation of this matrix
requires a product and four sums per sample, as well as a inversion
once the full matrix accumulated.  However, we can take advantage of
our prior knowledge of the problem to pre-compute the solution at a
given $T_{\rm int}$ and store it in RAM. For every possible
integration time up to $300 \,$s, we would only require 25.6 kbits to
store every possible matrix inverse as a lookup, which is negligible
compared to other RAM requirements.

For the $A^{\rm T} q$ term, we note again the structure of $A$, which
causes this term to simplify to:
\begin{equation}
 (A^{\rm T} q) = \begin{bmatrix} \sum_{0}^{n-1} A_{0i} q_{i} &
    \sum_{0}^{n-1} q_{i} \end{bmatrix}
\end{equation}
Of course, since $A_{0i} = i$ this simplifies further to:
\begin{equation}
 (A^{\rm T} q) = \begin{bmatrix} \sum_{0}^{n-1} i q_{i} &
    \sum_{0}^{n-1} q_{i} \end{bmatrix}
\end{equation}
To compute $ (A^{\rm T} q)$ in Equation \ref{eq:linefit}, we simply
need to compute two terms: the sum of the product of the frame number
and the data point at that frame, and the sum of the data points.  At
the end of the integration, we can multiply these by the pre-computed
$(A^{\rm T} A)^{-1}$ lookup to solve for $m$.

The resource requirements for this formalism are thus:
\begin{enumerate}
  \item The frame number $i$.  This can be shared by all fits as a
    global 8-bit integer.
  \item A cumulative sum of the product of $i$ and $q_{i}$ for each
    pixel, with size $4 \times 2048 \times 2048 \times 32$ bits.
  \item A cumulative sum of $q_{i}$ for each
    pixel, with size $4 \times 2048 \times 2048 \times 32$ bits.
\end{enumerate}
At the end of the integration, we multiply these cumulants by the
pre-computed inverse design matrix to compute
$\partial q / \partial t$ per pixel, requiring
$2 \times 4 \times 2048 \times 2048$ multiplies and
$4 \times 2048 \times 2048$ adds and resulting in a matrix with size
$4 \times 2048 \times 2048 \times 32$ bits.

\subsection{Transient Detection}

Though cosmic rays, electronics glitches, \textit{etc.} arise from
different physical processes, they all lead to step discontinuities in
the detector time streams.  These can be detected \textit{a
  posteriori} by determining goodness-of-fit statistics, but the
computational and memory burdens of such computations are large.  Here
we implement a ``transient detector'' algorithm that searches for step
discontinuities in real time and flags data points which exceed some
threshold from the expected linear trend for the detector.  The
resources required for this calculation are significantly less than
for a formal estimate of \textit{e.g.}  $\chi^{2}_{\nu}$.

A drawback of this algorithm is that it is less sensitive to
non-linear ramping effects that might affect the data.  Typically,
these effects occur in the same population of pixels in the array, so
ideally such behavior would be characterized and flagged during ground
testing.  The algorithm \emph{is} sensitive to non-linear ramps to a
certain extent defined by the detection threshold; the details of this
sensitivity will be characterized during ground testing so it can be
properly simulated in later analysis.

To detect transients, we can compute a residual function $\Delta r_{i}$:
\begin{equation}
  \label{eq:deltar}
  \Delta r_{i} = (q_{i} - \theta_{i})^{2}, 
\end{equation}
where $q_{i}$ is defined in Equation \ref{eq:qi} and $\theta_{i}$ is a
model for the expected value of the data at $i$ based on the previous
samples.  To flag transients, we require $\Delta r_{i}$ exceed some
pre-determined threshold $t$ times an estimate for the native variance
of the time stream $\sigma^{2}_{i}$:
\begin{equation}
  \label{eq:estimate}
  (q_{i} - \theta_{i})^{2} > t \sigma^{2}_{i}.
\end{equation}
The value of $t$ can be tuned using simulations or data (see
Sec.~\ref{ssS:crperformance}).  It is important that the estimate for
$\sigma_{i}^{2}$ be computed using a real-time estimate for the pixel
variance, since photon noise is dominant in each integration.  Since
each pixel will fall on a sky position with different brightness in
each integration, the photon noise will vary considerably, and a
single fixed variance estimate is not appropriate.

The problem to be solved therefore reduces to developing
suitable expressions for $\theta_{i}$ and $\sigma_{i}$.  In the
absence of noise, this is trivial, with
$\theta_{i} = i \cdot q_{i-1} / (i-1)$ and $\sigma^{2}_{i} = 0$.  However,
with noise in the time stream, an estimate for $\theta_{i}$ based
solely on $d_{i-1}$ will be unpredictable and reduce our sensitivity
to glitches.  Furthermore, because the charge naturally ramps, an useful
estimate for $\sigma^{2}$ needs to suppress gradients in $q(t)$.

We can form a more reliable estimate for $\theta_{i}$ by observing
that it should follow an equivalent expression for Equation
\ref{eq:qi}:
\begin{equation}
\theta_{i} = f \cdot i + \theta_{0}
\end{equation}
where $\theta_{0}$ is some initial value of $\theta$.  To compute this
expression, we require an estimate for $f$, which we can write as:
\begin{equation}
f_{i} = \frac{\sum_{j=0}^{i}(q_{j} - q_{0})}{\sum_{j=0}^{i} (j)}
\end{equation}
If we substitute this expression straight into Equation
\ref{eq:deltar}, all terms will cancel.  Instead, we use the sum of the
$(i-1)$ measurements to estimate the slope, thereby reducing the error
by the square root of the number of samples.  That leads us to:
\begin{equation}
  \theta_{i} = \left( \frac{\sum_{j=0}^{i-1}(q_{j}) - (i-1)
    q_{0}}{\sum_{j=0}^{i-1}(j)} \right) \cdot i + \theta_{0},
\end{equation}  
and as a result:
\begin{equation}
  \Delta r_{i} = (q_{i} - \theta_{i})^{2} = \left( q_{i} -
  \left\{ \left(\frac{i}{i-1}
  \right) \cdot (\sum_{j=0}^{i-1}(q_{j}) - (i-1) q_{0}) + \theta_{0} \right\}
  \right)^{2} .
\end{equation}
A simplifying choice is to set $\theta_{0}=q_{0}$ measured for that pixel.

Computationally, this algorithm requires the storage of:
\begin{enumerate}
\item $q_{i}$, which we already have stored from the line fit;
\item $i$, which we also already have from the line fit;
\item the sum of the previous samples $\sum_{j=0}^{i-1} q_{j} $, which we
  already have from the term $\sum q_{i}$ from the line fit, if
  we do the transient computation \emph{before} adding $q_{i}$ to
  $q_{i-1}$; and
\item the initial read value $q_{0}$.  This is the only new memory
  requirement and would have size $4 \times 2048 \times 2048 \times
  32$ bits.
\end{enumerate}
In the worst case, the computation requires:
\begin{enumerate}
\item a product of two numbers $(i-1)$ and $q_{0}$;
\item a difference between $\sum q_{i-1}$ and the output of step 1;
\item a product and divide of the number computed in step 2 with $i/(i-1)$;
\item a sum of the output of step 3 with $q_{0}$;
\item a difference between $q_{i}$ and $\theta_{i}$ computed in step 4;
\item a product of the output of step 5; and
\item a comparison operation between the output of the calculation and
  the product of a lookup value loaded in memory and $\sigma_{i}^{2}$.
\end{enumerate}

To develop an expression for $\sigma_{i}^{2}$, we observe that the
optimal estimate for the variance of a set of numbers $q_{i}$ is given
by
\begin{equation}
\sigma^{2} = \frac{1}{n-1} \sum_{i=0}^{n} (q_{i} - \bar{q})^{2}
\end{equation}
with $n$ as the number of samples in the set, and $\bar{q}$ the mean
of the set.  In the presence of a gradient (as in Equation \ref{eq:qi}), this can be
generalized to:
\begin{equation}
 \sigma^{2} = \frac{1}{n-1} \sum_{i=0}^{n} (q_{i} - (f \cdot i + q_{0}))^{2}.
\end{equation}
The term in the parenthesis is similar to our residual
estimate $\Delta r_{i}$. As a result:
\begin{equation}
  \sigma^{2}_{i} \approx \frac{1}{i-1} \sum^{i}_{j=0}
  \Delta r_{j} 
\end{equation}
We can then note that:
\begin{equation}
  \sigma^{2}_{i} = \frac{1}{i-1} \left( \Delta r_{i} + (i-2)
  \sum_{j}^{i-1}\Delta r_{j} \right).
\end{equation}
This estimate requires the storage of $4 \times 2048 \times
2048 \times 32$ bits.  Since $\Delta r_{i}$ is computed at each step anyway,
the additional computational requirements are modest, namely:
\begin{enumerate}
\item a product of $i-2$ and the running sum of residuals $\sum^{i-1} \Delta
  r_{j}$;
\item a sum of $\Delta r_{i}$ and the output of step 1; and
\item a division of the output of step 2 by $i-1$.
\end{enumerate}

Note that, though what we term $\sigma_{i}^{2}$ is related to the
variance of the data, in fact it does not reproduce the variance
calculation in an individual measurement.  This is because it is
formed from a running estimate of the mean, which is sensitive to
deviations caused by noise.

\subsection{Detector Saturation}

Charge-integrating detectors have a finite well capacity, and due to
mutual repulsion of electrons, also suffer from reciprocity failure at
large integrated charge (\textit{e.g.} \citealt{Biesiadzinski2011}).
It is very expensive to correct this effect in real-time, since each
pixel will have its own reciprocity behavior as a function of
integrated charge; correction entails building a lookup table for each
pixel and computing a correction factor which can be an arbitrarily
complex function of $q$.  To account for this behavior, we flag and
ignore further samples for those pixels whose accumulated charge
exceeds some preset threshold $q_{\rm max}$, so pixels are flagged
when:
\begin{equation*}
q_{i} - q_{0} > q_{\rm max}
\end{equation*}
Since these quantities are stored for the photocurrent estimator, this
requires no new storage requirements except for the single, global
32-bit threshold value.  The value of $q_{\rm max}$ can be determined
pre-flight during laboratory characterization of each detector.  This
step does require two additional computations:
\begin{enumerate}
\item a difference of $q_{i}$ and $q_{0}$;
\item a comparison operation between the output of the calculation and
  $q_{\rm \max}$.
\end{enumerate}
Accurate photocurrent estimates can still be generated for these
pixels, as discussed in Sec.~\ref{sS:truncatedfits}.

\subsection{Early Reads}
\label{sS:earlyreads}

HgCdTe arrays are known to have unpredictable behavior in the first few
reads after a reset due to charge injection and other transient
effects.  As a result, it is desirable to include capability to ignore
the first few frames in a fit (likely values run anywhere from 0 to
5).  The precise number will be determined during instrument
characterization; in this work, we assume a skip value of 3 frames;
the results we present are not sensitive to this choice.

In addition, we have the problem that, at early frame numbers, there
is not yet enough information accumulated to produce a reliable
estimate for $\sigma^{2}_{i}$.  Fortunately, if the first few reads
are ignored from the fit, we can use them to generate an estimate for
$\sigma^{2}_{i}$ based on the simple difference of $q_{i}$ and
$q_{0}$, namely:
\begin{equation}
\Delta r_{i} = (q_{i} - q_{0})^{2}.
\end{equation}
This does not impose any additional storage or computational
requirements on the algorithm. 

The threshold $t$ in Equation \ref{eq:estimate} governs the effective
number of standard deviations at which a transient event is detected.
In practice, we find that the algorithm is more sensitive to steps
later in the time series because the variance estimate needs to be
``trained'' to reliably find steps.  We empirically find that the
training precision improves linearly with time, so we modify Equation
\ref{eq:estimate} to include a weighting by the frame number as:
\begin{equation}
  \label{eq:newestimate}
  \Delta r_{i} = (d_{i} - m_{i})^{2} > \frac{t \sigma^{2}_{i}}{i}.
\end{equation}
This modification effectively flattens the glitch detection
efficiency.  The value of $t$ is optimized using simulations of the
system (see Sec.~\ref{ssS:crperformance}).

\subsection{Truncated Fits}
\label{sS:truncatedfits}

For pixels suffering from cosmic ray events and saturation, it is
possible to form a photocurrent estimate using the data acquired
before the pixel data are flagged.  In this case, we simply fit to the
pre-flag data as input to the algorithm described above, using a
solution to $(A^{\rm T} A)^{-1}$ appropriate for the length of data.
These will be flagged as suspect in the telemetered data, and can be
incorporated (or not) into later analysis as appropriate.  For
SPHEREx, this will allow retrieval of a photocurrent estimate for all
pixels falling on bright stars with magnitudes $J > 8$, which occur at
a rate of $\sim 5 \,$per square degree in high galactic latitude
fields.

\section{Algorithm Characterization}
\label{S:testing}

Having developed the algorithm, we can test its performance
in realistic scenarios.  

\subsection{Input Simulations}
\label{sS:input}

The input data simulation comprises three components.  First, we
simulate H2RG noise with the {\sc nghxrg} package, which simulates
significant noise components including: white read noise; residual
bias drifts; pink $1/f$ noise; alternating column noise; and picture
frame noise \citep{Rauscher2015}. In the simulation, we specify
32-output mode driven by a SIDECAR ASIC; the parameters of the
simulation are specified in Table \ref{tab:arrays}.  Second, we
simulate dark current appropriate for a typical H2RG.  The current is
simulated as a random draw from a Poisson distribution with mean
$i_{\rm d}$.  This accurately simulates the effect of random thermal
electron generation in the detectors.  Finally, we simulate a
flight-like photon load by generating an astrophysical image including
emission from ZL, diffuse galactic light, stars, and galaxies. These
components are computed in surface brightness units, and are then
calibrated to photocurrent $i_{\rm phot}$ using Equation \ref{eq:Iph}.
An example input astronomical image is shown in Fig.~\ref{fig:output}.
Photon noise is accounted for in the same way as for dark current,
with a random draw from a Poisson distribution with mean
$i_{\rm phot}$.  We then add cosmic rays to this simulation at a rate
and with an energy spectrum determined from HST-WFC3 data as discussed
in Sec.~\ref{sS:cosmicrays}.  We use statistical realizations of
the measured mean cosmic ray spectrum for regular geomagnetic regions,
though the algorithm performance scales to regions with increased event
rates as expected.

\subsection{Algorithm Performance}

To verify the performance of the algorithm, we use the input simulated
read data, propagate it through the steps described in
Sec.~\ref{S:dataprocessing}, and compare the output both to the input
and to the output of a standard line fitting package.

\subsubsection{Photocurrent Estimates}
\label{ssS:photocurrent}

\begin{figure*}[ht]
\centering
\includegraphics*[width=6.5in]{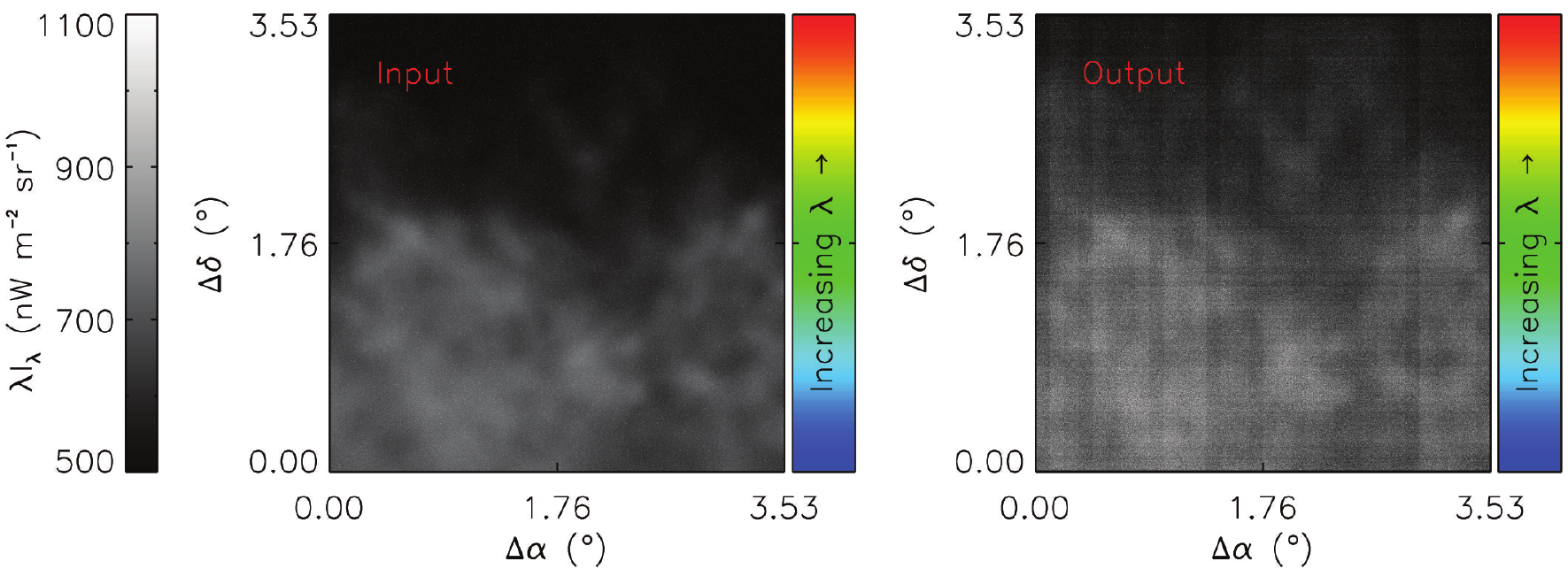}
\caption{Simulated SPHEREx images, both for input and after processing
  by the on-board photocurrent algorithm.  (\textit{LEFT}) An input
  sky image including emission from Zodiacal light, diffuse galactic
  light, stars, and galaxies for SPHEREx band 1.  The image is
  presented as a surface brightness $\lambda I_{\lambda}$, so that the
  decreasing brightness of Zodiacal light with increasing wavelength
  (as indicated by the bar) causes a visible gradient in the image.
  (\textit{RIGHT}) Output photocurrent estimate following the addition
  of instrument noise, cosmic ray events, and processing by the
  on-board photocurrent algorithm, again expressed as
  $\lambda I_{\lambda}$.  The banding structure visible in the image
  is due to uncompensated correlated noise in the H2RG simulation.
  The 32 readout channels run vertically, so noise variations common
  to all output channels show up as horizontal stripes.  We plan to
  compensate for these using an electronic referencing scheme similar
  to that presented in \citet{Moseley2010} in the flight
  system that successfully reduces the correlated noise. \label{fig:output} }
\end{figure*}

An image of both an input sky image, and the output following the
on-board processing simulation is shown in Fig.~\ref{fig:output}.
We find that the photocurrents are unbiased, in the sense that in the
absence of instrument noise they return the input image to floating
point error.  Likewise, in pixels with no simulated cosmic ray events,
both our algorithm and the output of a standard fitting
package\footnote{\url{http://docs.scipy.org/doc/numpy-1.10.0/reference/generated/numpy.polyfit.html}}
return the same values to floating point error.  This satisfies
requirement 1 listed in Sec.~\ref{S:dataprocessing}, since there is
no discernable statistical difference between the performance of this
algorithm and that of a standard line fitting algorithm under the same
noise assumptions.

The assumption of constant noise required to simplify the algorithm is
violated with these charge-integrating detectors since the photon
noise grows in each read.  However, we have tested that this algorithm
still returns an unbiased estimate of the photocurrent using our
simulations, and find no evidence for bias.  This is due to the fact
that, even though the reads toward the end of the integration are
overweighted, they are not themselves biased compared to the model.
If we were to use this algorithm to derive optimal fit parameter
uncertainties, $\chi^{2}$, or other fit statistics sensitive to the
instantaneous noise estimate we would find a significant underestimate.

\subsubsection{Cosmic Ray Rejection Performance}
\label{ssS:crperformance}

\begin{figure*}[ht!]
\centering
\includegraphics*[width=6.5in]{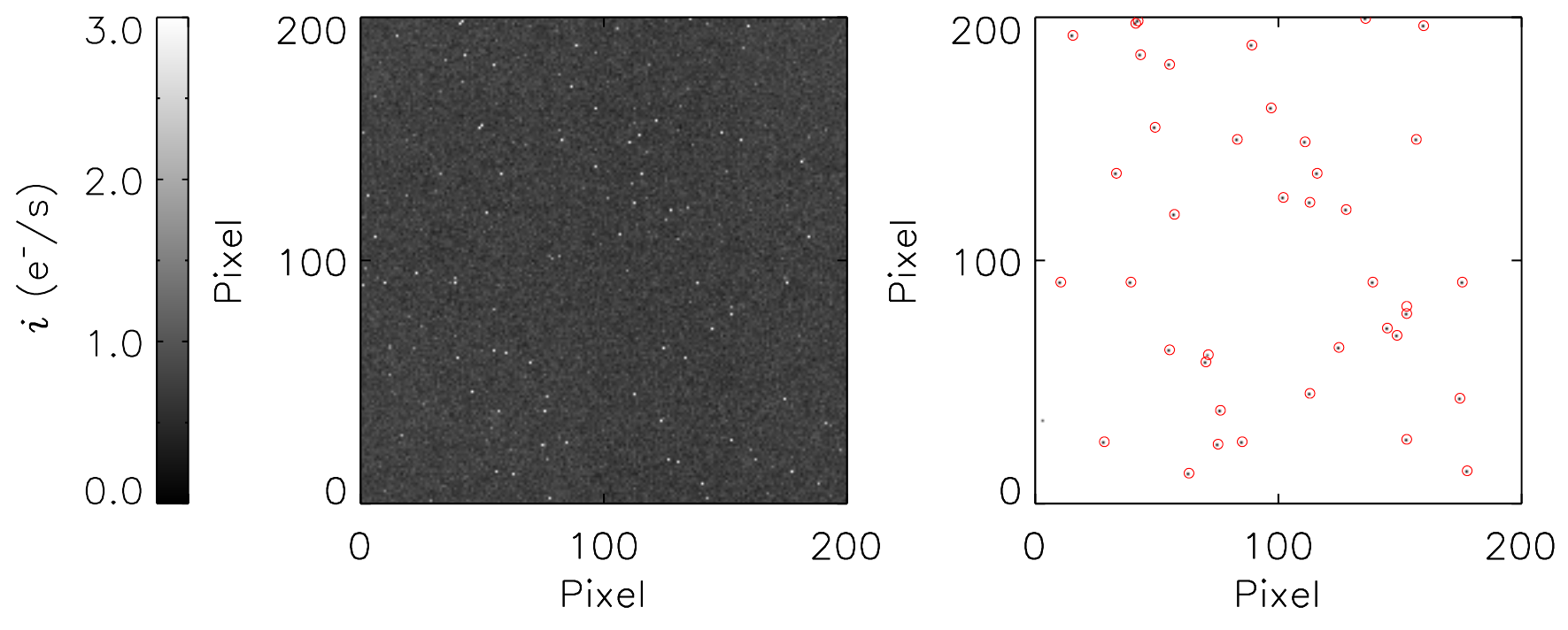}
\caption{Time series transient rejection visualization.
  (\textit{LEFT}) A $200 \times 200$ pixel sub-image of the simulation
  output (analogous to the right hand panel in Fig.~\ref{fig:output})
  with no cosmic ray rejection applied.  The bright pixels are due to
  both cosmic rays and bright stars and galaxies in the image.
  (\textit{RIGHT}) The grey scale image shows the difference between
  photocurrent estimates with and without transient detection; cosmic
  rays that are found with transient detection appear as black dots.
  The population of input cosmic rays are also shown (red circles).
  Of the 40 transient events in this sub-image, 39 are found
  successfully, and the missed event (near \{5,40\}) has low energy.
  One event is flagged with no corresponding input cosmic ray event
  input (near \{155,80\}).  \label{fig:tfsf} }
\end{figure*}

To determine the efficacy of the cosmic ray rejection algorithm, we
have measured several metrics.  The simplest of these is to check that
images with and without cosmic ray rejection include approximately 500
fewer large-value pixels.  In Fig.~\ref{fig:tfsf}, we show the
difference between a single image realization with and without cosmic
ray rejection for a $200 \times 200$ pixel sub-image.  The cosmic
rays are detected with high efficiency.

\begin{figure*}[ht!]
\centering
\includegraphics*[width=6.5in]{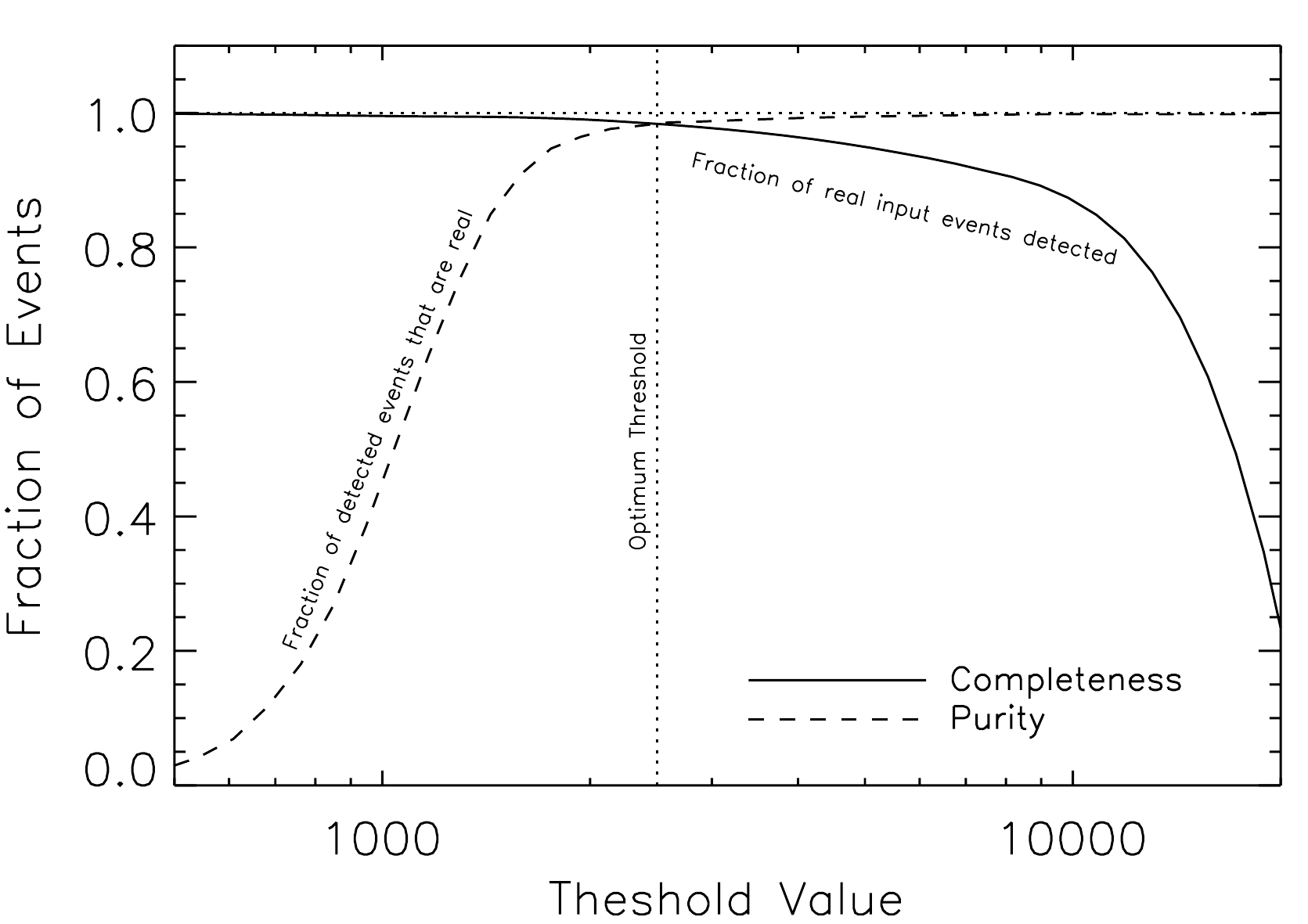}
\caption{The purity and completeness of the cosmic ray detection
  algorithm as a function of the detection threshold $t$, optimized
  at $t_{\rm max}=2500$ for a simulated $T_{\rm int} = 107 \,$s
  SPHEREx observation.  \label{fig:pc} }
\end{figure*}

To maximize the performance of the algorithm, the parameter $t$ must
be optimized.  We optimize against two metrics:
\begin{itemize}

  \item \emph{completeness}, defined as the ratio of detected events
    to the total number of injected cosmic rays; and

  \item \emph{purity}, defined as the ratio of detected cosmic rays that
    were injected to all detected events.

\end{itemize}
In Fig.~\ref{fig:pc} we show the purity and completeness of the
cosmic ray detection as a function of $t$.  As expected, purity and
completeness are anti-correlated.  This can be understood by the fact
that, as we reduce the value of $t$, more candidate events will be
flagged, which increases both the number of detected input events as
well as the number of false detections.  If we increase the value of
$t$, the total number of detections will be reduced, which increases
the number of false detections, but simultaneously missing more input
cosmic rays.  We find an optimum $t_{\rm max} = 2500$, where the
purity is $>0.993$, and the completeness is $>0.985$.

\begin{figure*}[ht!]
\centering
\includegraphics*[width=6.5in]{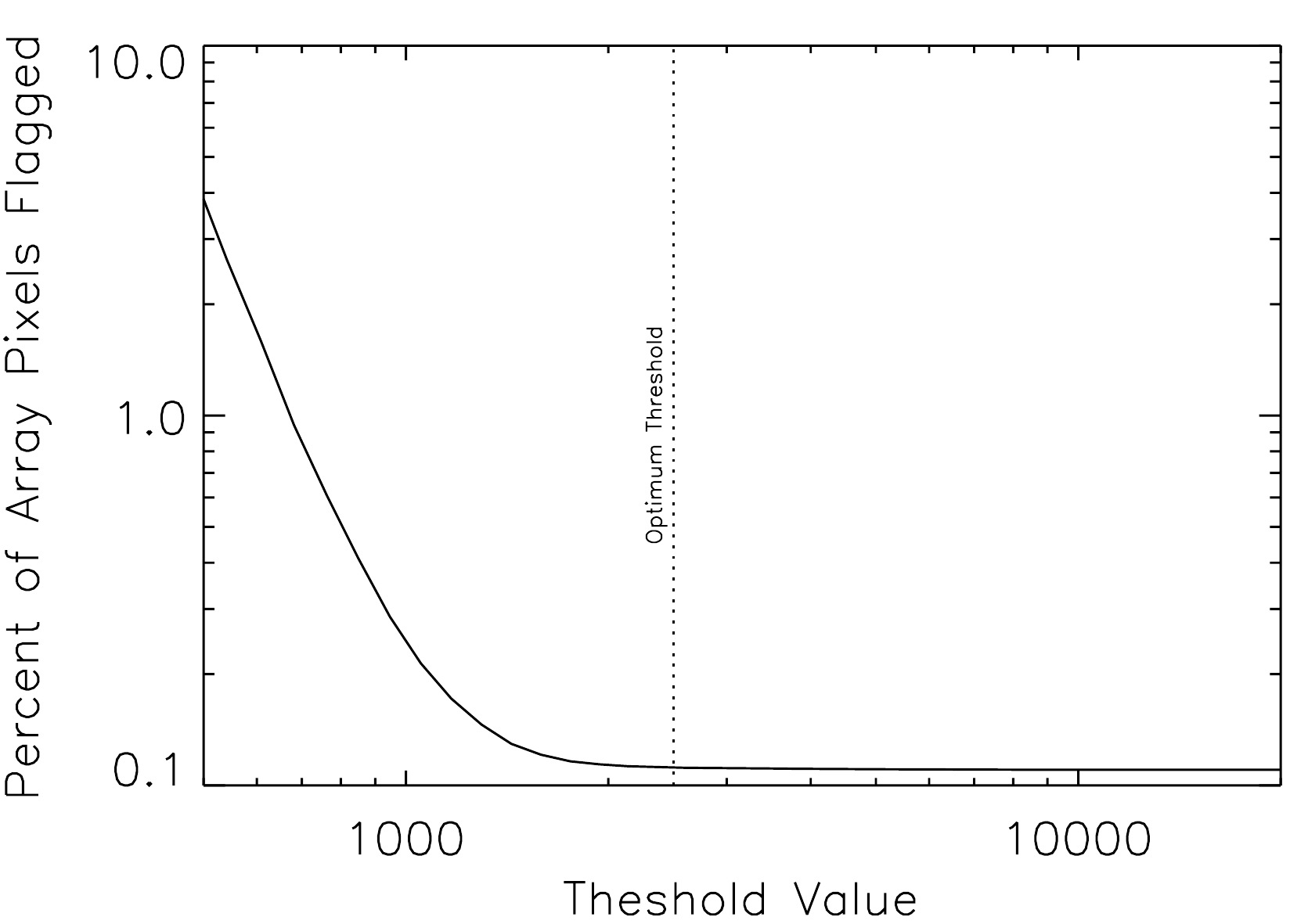}
\caption{Pixel flagging fraction as a function of $t$.  At
  $t_{\rm max}$ we predict a pixel loss fraction of $\sim 0.11$\% in a
  single integration.  \label{fig:typ} }
\end{figure*}

The pixel loss fraction can be calculated from the purity and
completeness.  As shown in Fig.~\ref{fig:typ}, as $t$ is varied, the
number of flagged pixels changes, with a larger loss fraction (more
flagged pixels) at smaller values of $t$.  At $t_{\rm max}$, we
predict a pixel loss fraction of $\sim 0.11$\%, which
  has a negligible impact on SPHEREx science \citep{Dore2014}.

\begin{figure*}[ht!]
\centering
\includegraphics*[width=6.5in]{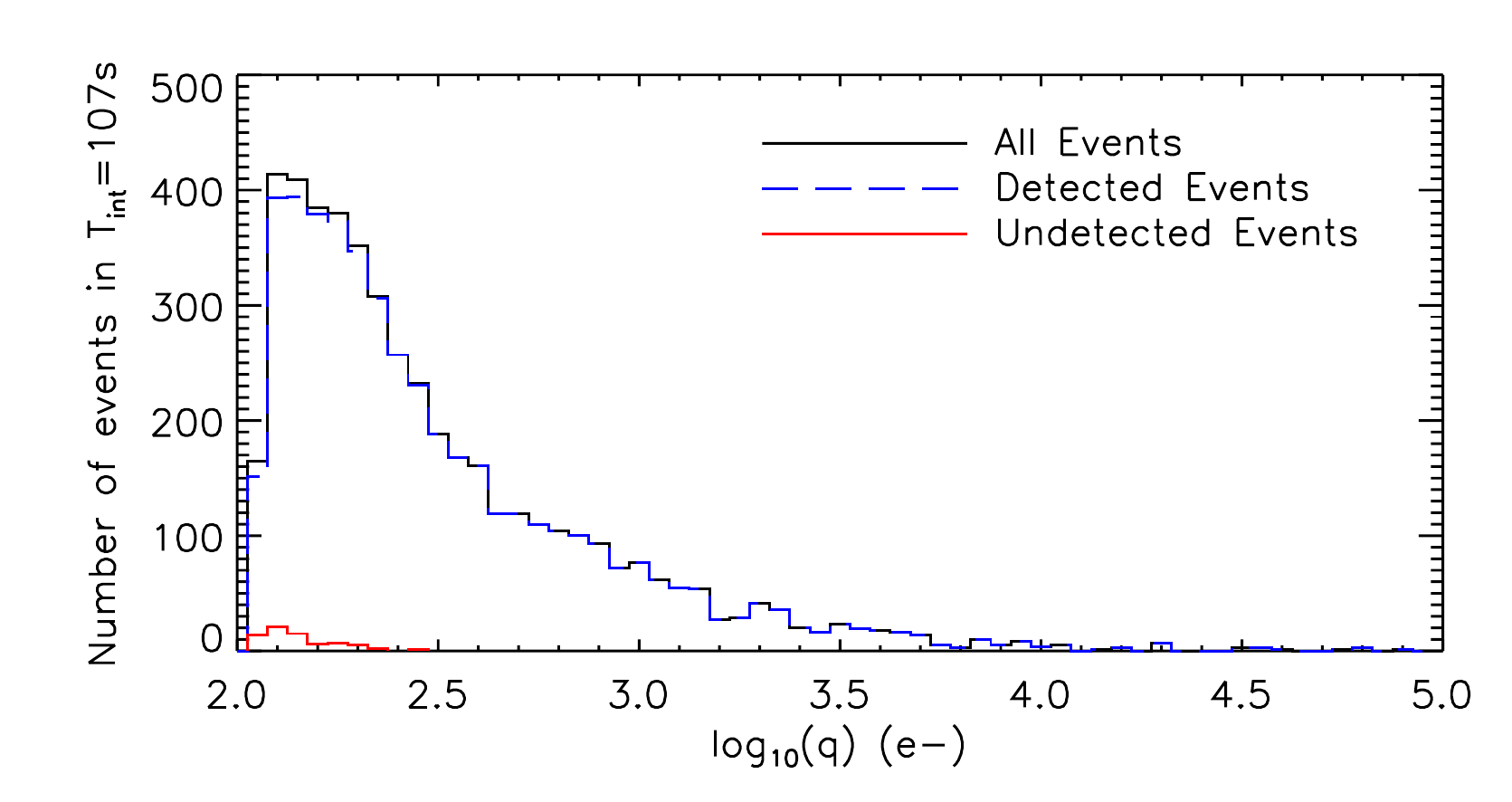} \\
\includegraphics*[width=6.5in]{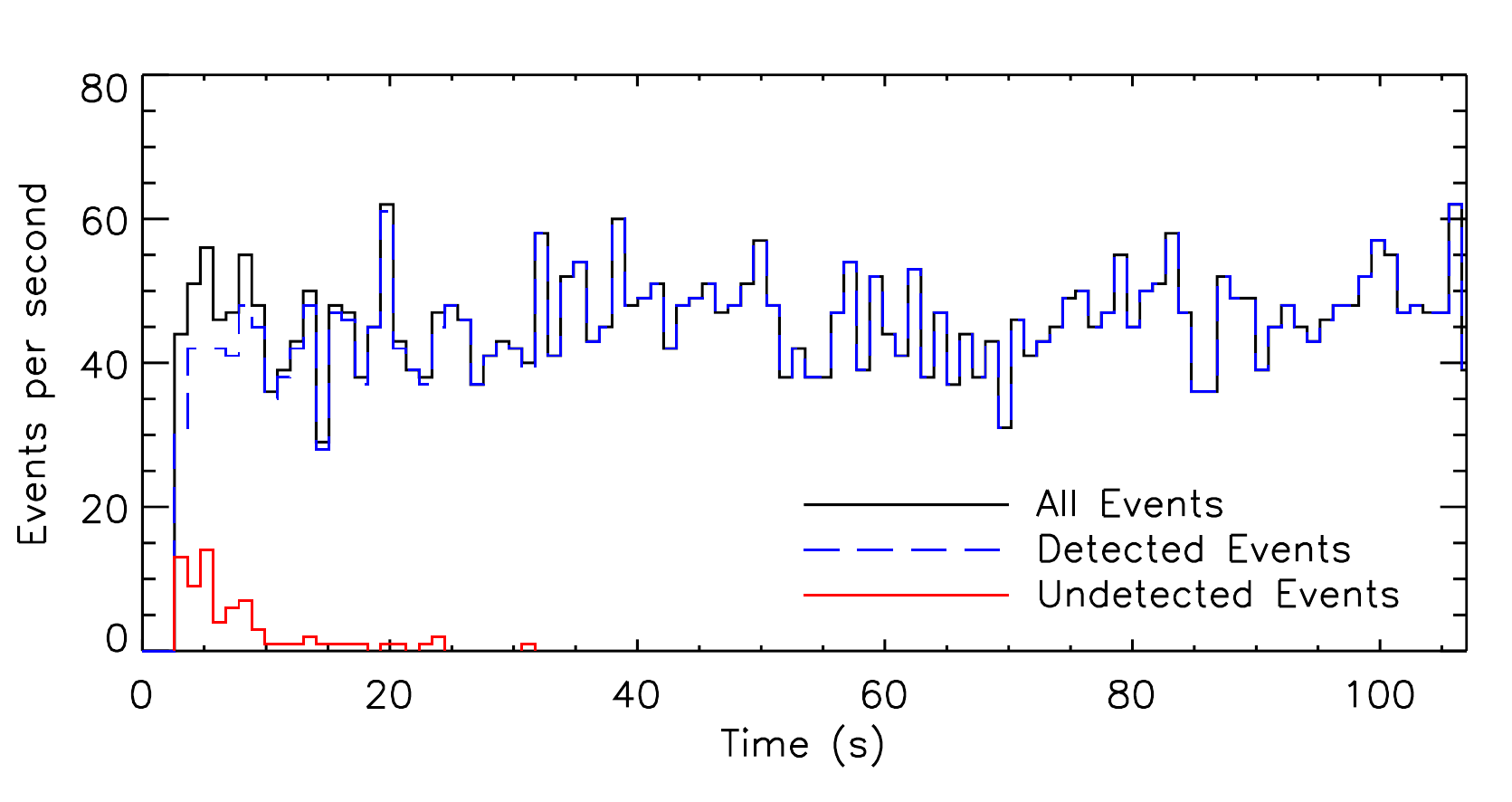} 
\caption{Cosmic ray energy and timing diagnostics for all, detected,
  and undetected events.  (\textit{TOP}) Histogram of cosmic ray
  deposition energies in a single simulated $T_{\rm int} = 107 \,$s
  integration.  We also indicate the population of detected events,
  and undetected events as indicated in the plot.  In all, $98.5 \,$\%
  of all cosmic ray events are detected.  The undetected events lie at
  low energy where they are difficult to detect against read noise.
  (\textit{BOTTOM}) Cosmic ray event times in a single simulated
  $T_{\rm int} = 107 \,$s integration.  We indicate the population of
  all events, the detected events, and the undetected events as shown
  in the legend.  Undetected events lie at early times when the error
  in the variance estimate used in the step detection is large.  This
  is an unavoidable consequence of using a real-time variance
  estimator.\label{fig:fc1} }
\end{figure*}

Fixing to $t=t_{\rm max}$, we can investigate the performance of the
cosmic ray detection algorithm versus: (\textit{i}) cosmic ray energy;
and (\textit{ii}) time in the integration.  In Fig.~\ref{fig:fc1} we
show a histogram of the event energies in a single simulation.
Undetected cosmic ray events occur at low energies where they are more
difficult to detect against the detector noise.  The detection
probability as a function of time is illustrated in
Fig.~\ref{fig:fc1}.  The algorithm is more likely to miss events close
to the beginning of the integration because the running variance
estimate has a large uncertainty during early reads (see
Sec.~\ref{sS:earlyreads}), so cosmic ray outliers have a greater
chance of passing as a normal statistical variation.

SPHEREx plans to implement both the standard $107 \,$s integration
time, as well as a deeper $173 \,$s integration for use in ``deep
field'' regions.  We have simulated these longer integrations and find
similar behavior to the standard length exposures.  The absolute count
of undetected cosmic ray events is the same, since at early reads
where cosmic rays are typically missed the algorithm performance is
independent of integration time.  The completeness and purity of the
longer integration times expressed as a fraction of total events
improves since the late-time performance of the algorithm is close to
ideal.

The probability of our algorithm detecting cosmic rays is largest at
late times and large energies.  As a result, we expect undetected
cosmic rays to manifest as a slight positive-going non-Gaussianity in
the flight image noise properties.  Accounting for this is best
handled through detailed simulation of the on-board processing using
the measured flight cosmic ray rate and spectrum in the SPHEREx flight
data.  The resulting image simulations will need to be propagated
through the full SPHEREx data reduction pipeline to build a detailed
understanding of their effect on the science results (see
\textit{e.g.} \citealt{Zemcov2014} for an example of such an analysis).

\section{Hardware Implementation}
\label{S:hardware}

The SPHEREx flight hardware provides a platform to perform on-board
data analysis and reduction, as well as long-term storage prior to
telemetry to the ground station.  

\subsection{Flight Processing Capabilities}
\label{sS:flightprocessing}

The SPHEREx warm electronics uses a digital board with a Virtex 5,
including 1GB SDRAM, and 128GB Flash memory for on-board data
processing. The Virtex 5 performs all calculations needed for the SUR
algorithm. The SDRAM provides storage of the partial sums and flag
information during the exposure, and also has separate long-term
storage for the final image data. Key capabilities of the Virtex 5 are
20480 Slices, 320 DSP-48s, 1.34 MB embedded RAM, and a 550MHz max
clock
frequency\footnote{\url{http://www.xilinx.com/onlinestore/silicon/online_store_v5.htm}}.
The warm electronics includes an additional RAD750 computer. This is
used solely to compress the final images before downlink, and does not
participate in the SUR calculation.
 
\subsection{Implementation Tests}
\label{sS:implementation}

\begin{figure*}[!p]
\centering
\includegraphics*[width=6.5in]{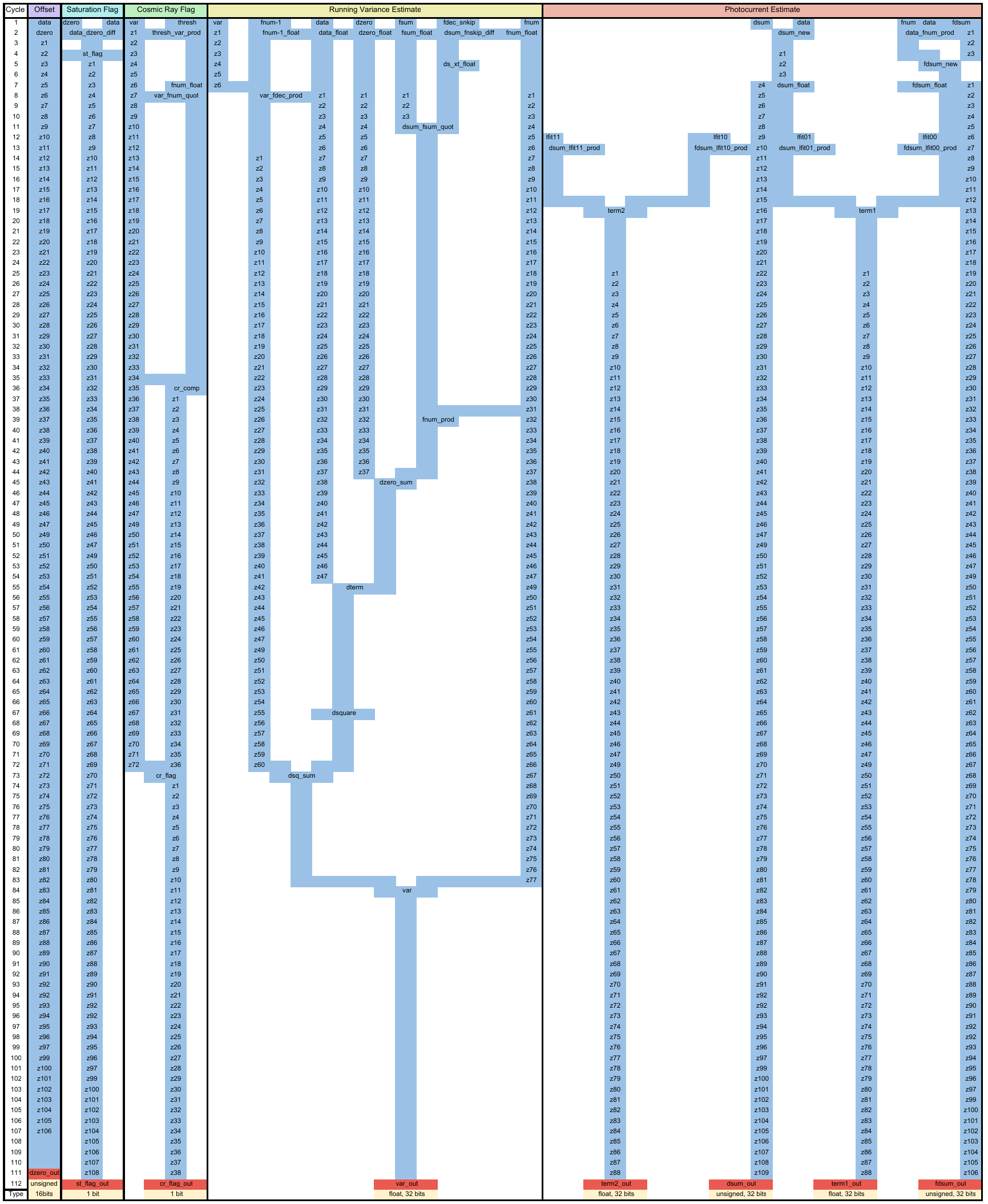} 
\caption{A schematic of our on-board processing algorithm as implemented in a
  hardware simulation.  Each row represents one $100 \,$MHz clock
  cycle.  Cells containing variable names show the time required to
  produce an intermediary product of the algorithm.  The number of
  cycles required to complete an operation depends on the complexity
  of the operation.  Cells containing a "zX" value represent delays
  wait states required to ensure that input operand values are
  available during computation.  The running variance calculation
  requires the largest numbers of cycles to complete, largely because
  of the multiple floating point division operations in the
  calculation.  \label{fig:hw} }
\end{figure*}

It is standard practice to program FPGA code and simulate its
implementation in FPGA on a desktop computer. We have performed this
exercise with the algorithm presented in Sec.~\ref{S:dataprocessing}
using Xilinx ISE 14.7 simulating a Virtex 5 XC5VFX130T device.  The
algorithm was reduced to its elementary operations and a pipeline was
developed to complete the necessary calculations in hardware, as shown
schematically in Fig.~\ref{fig:hw}.  Xilinx Coregen was used to
generate p-cores that performed the arithmetic.  Shift registers and
control logic were added to complete the design. Modelsim was used as
the primary debugging tool throughout the development process.  A
sample data stream was presented to the hardware and software versions
of the algorithm and both implementations returned the same slope and
flag data. Because of this, we are confident that resource allocations
are representative of those required in a hardware implementation of
the algorithm.

\afterpage{\clearpage}

\subsubsection{FPGA Utilization}
\label{ssS:fpgaimplimentation}

The FPGA performs calculations and operations by manipulating
configurable logic blocks and utilizing on board digital signal
processors (DSPs).  The Virtex-5 DSP48 slices are the portions of the
FPGA specifically designed for performing
arithmetic. The Look-Up Table (LUT) metric represents
  combinational logic usage and the Flip-Flop (FF) metric represents
  sequential logic usage \citep{Virtex5}. The as-implemented resource
utilization of the SPHEREx digital board is shown in Table
\ref{tab:utilization}. The FPGA utilization is at $12\%$ or below in
all respects, meaning there is ample margin, and there is room to add
additional capabilities in further studies.  In addition,
post-synthesis analysis reported a minimum clock period of 4.390 ns.
This corresponds to a maximum clock frequency of 227 MHz which easily
meets the requirements of the design.

\subsubsection{Memory Interface Utilization}
\label{ssS:memaimplimentation}

Part of the main processing engine is an interface to the external
memory. During the processing of the nearly 17 million pixels, each of
the partial sums from Sec.~\ref{S:dataprocessing} needs to read,
manipulated, and then stored in memory every 1.5 seconds.  The
processing platforms must have the resources to store this data, and
have bandwidth to read and write this data once every 1.5 seconds. The
storage necessary to hold the partial sum, flag data, and leave space
for the final image is 275 MB, and our capacity is considerably larger
at 1 GB. The requirement on the interface is to read and write the
partial sum and flag data, and yields a required rate of 320 MB per
second. We compare this rate to a theoretical capacity of our
SDRAM. Using 32-bit DDR SDRAM at 200 MHz with an assumed $67\%$
bandwidth efficiency gives a capability of 1067 MB per second. This
capability of 1067 MB per second is again considerably more than the
required 320 MB per second.

\begin{table}
\small
\centering
\caption{FPGA and memory utilization by the SUR algorithm. \label{tab:utilization}}
\begin{tabular}{lccccc}
\hline
Resource & Usage & Available & Utilization $[\%]$  \\ \hline

DSP  & 34 & 320 &11 \\ 
LUTs & 7497 & 81920 & 9 \\ 
FFs & 9746 & 81920 & 12 \\ 
SDRAM & 275 & 1000 & 28 \\ 
Mem I/F [MB/s] & 320 & 1067 & 30 \\ \hline

\end{tabular}
\end{table}

\section{Discussion}
\label{S:discussion}

In this paper we provide a practical example of an on-board data
processing and reduction algorithm that can fit in a
resource-constrained system.  Implementation of optimal photocurrent
estimators are useful in similar resource-constrained systems, a
salient example being CubeSats for astrophysics.  Some type of
on-board processing and data compression is required to fit into the
limited RAM, processing capability, long-term storage, and telemetry
bandwidths available in such small platforms.  Our algorithm is a
low-footprint, fast, hardware-demonstrated solution well suited to the
needs of charge-integrating detectors, which are very common from the
UV to mid-IR wavelengths.  It successfully meets accuracy metrics for
an optimal photocurrent estimate, and has been demonstrated in a
system with flight heritage.

We have demonstrated that processing platforms with significant flight
heritage are significantly under-utilized by our algorithm.  This
provides space for other data acquisition tasks, if required, as well
as showing this algorithm could fit on even more constrained
platforms, if necessary.  In the particular case of
  SPHEREx, we note there remain significant resources available to
  implement some form of correlated noise reduction, though part or
  all of the requirements for such a scheme may fall on \textit{e.g.}
  the ASIC.

The algorithm discussed here has applicability beyond astrophysics, in
applications as diverse as earth sensing, planetary studies, attitude
control, surveillance, or any resource-constrained situation where
optimal photocurrent estimates are required and time-space transients
are common.

\section*{Acknowledgments}

Our sincere thanks to S.~Baggett and E.~Wright, who graciously sent
HST-WFC3 and WISE cosmic ray data (respectively), and helped us
understand how to interpret them.  We would like to acknowledge the
use of the {\sc nghxrg} software package written by B.~Rauscher, and
thank him for making such a valuable tool public.

This research was funded, in part, by the NASA Astrophysics Explorer
Program.  Part of this research was carried out at the Jet Propulsion
Laboratory, California Institute of Technology, under a contract with
the National Aeronautics and Space Administration.  This publication
makes use of data products from the Wide-field Infrared Survey
Explorer, which is a joint project of the University of California,
Los Angeles, and the Jet Propulsion Laboratory/California Institute of
Technology, funded by the National Aeronautics and Space
Administration, and from the NASA/ESA Hubble Space Telescope, operated
by the Association of Universities for Research in Astronomy,
Inc. under NASA contract NAS 5-26555.

\bibliography{apj-jour,level0_simulation}

\begin{thebibliography}{23}
\newcommand{\enquote}[1]{``#1''}
\providecommand{\natexlab}[1]{#1}
\providecommand{\url}[1]{\texttt{#1}}
\providecommand{\urlprefix}{URL }
\expandafter\ifx\csname urlstyle\endcsname\relax
  \providecommand{\doi}[1]{doi:\discretionary{}{}{}#1}\else
  \providecommand{\doi}{doi:\discretionary{}{}{}\begingroup
  \urlstyle{rm}\Url}\fi

\bibitem[{{Barker} \emph{et~al.}(2010){Barker}, {McCullough} \&
  {Martel}}]{Barker2010}
{Barker}, E.~A., {McCullough}, P. \& {Martel}, A.~R. [2010]  \enquote{{WFC3 IR
  SAA Passage Behavior},} Tech. rep.,
  \urlprefix\url{http://www.stsci.edu/hst/wfc3/documents/ISRs/WFC3-2009-40.pdf}.

\bibitem[{{Beletic} \emph{et~al.}(2008){Beletic}, {Blank}, {Gulbransen}, {Lee},
  {Loose}, {Piquette}, {Sprafke}, {Tennant}, {Zandian} \& {Zino}}]{Beletic2008}
{Beletic}, J.~W., {Blank}, R., {Gulbransen}, D., {Lee}, D., {Loose}, M.,
  {Piquette}, E.~C., {Sprafke}, T., {Tennant}, W.~E., {Zandian}, M. \& {Zino},
  J. [2008]  \enquote{{Teledyne Imaging Sensors: infrared imaging technologies
  for astronomy and civil space},}  \emph{High Energy, Optical, and Infrared
  Detectors for Astronomy III}, p. 70210H, \doi{10.1117/12.790382}.

\bibitem[{{Biesiadzinski} \emph{et~al.}(2011){Biesiadzinski}, {Lorenzon},
  {Newman}, {Schubnell}, {Tarl{\'e}} \& {Weaverdyck}}]{Biesiadzinski2011}
{Biesiadzinski}, T., {Lorenzon}, W., {Newman}, R., {Schubnell}, M.,
  {Tarl{\'e}}, G. \& {Weaverdyck}, C. [2011]  \emph{\pasp} \textbf{123},  179,
  \doi{10.1086/658282}.

\bibitem[{{Bock} \emph{et~al.}(2013){Bock}, {Sullivan}, {Arai}, {Battle},
  {Cooray}, {Hristov}, {Keating}, {Kim}, {Lam}, {Lee}, {Levenson}, {Mason},
  {Matsumoto}, {Matsuura}, {Mitchell-Wynne}, {Nam}, {Renbarger}, {Smidt},
  {Suzuki}, {Tsumura}, {Wada} \& {Zemcov}}]{Bock2013}
{Bock}, J., {Sullivan}, I., {Arai}, T., {Battle}, J., {Cooray}, A., {Hristov},
  V., {Keating}, B., {Kim}, M.~G., {Lam}, A.~C., {Lee}, D.~H., {Levenson},
  L.~R., {Mason}, P., {Matsumoto}, T., {Matsuura}, S., {Mitchell-Wynne}, K.,
  {Nam}, U.~W., {Renbarger}, T., {Smidt}, J., {Suzuki}, K., {Tsumura}, K.,
  {Wada}, T. \& {Zemcov}, M. [2013]  \emph{\apjs} \textbf{207}, 32,
  \doi{10.1088/0067-0049/207/2/32}.

\bibitem[{{Crouzet} \emph{et~al.}(2012){Crouzet}, {ter Haar}, {de Wit},
  {Beaufort}, {Butler}, {Smit}, {van der Luijt} \& {Martin}}]{Crouzet2012}
{Crouzet}, P.-E., {ter Haar}, J., {de Wit}, F., {Beaufort}, T., {Butler}, B.,
  {Smit}, H., {van der Luijt}, C. \& {Martin}, D. [2012]
  \enquote{{Characterization of HAWAII-2RG detector and SIDECAR ASIC for the
  Euclid mission at ESA},}  \emph{High Energy, Optical, and Infrared Detectors
  for Astronomy V}, p. 84531R, \doi{10.1117/12.924968}.

\bibitem[{{Dor{\'e}} \emph{et~al.}(2014){Dor{\'e}}, {Bock}, {Ashby}, {Capak},
  {Cooray}, {de Putter}, {Eifler}, {Flagey}, {Gong}, {Habib}, {Heitmann},
  {Hirata}, {Jeong}, {Katti}, {Korngut}, {Krause}, {Lee}, {Masters},
  {Mauskopf}, {Melnick}, {Mennesson}, {Nguyen}, {{\"O}berg}, {Pullen},
  {Raccanelli}, {Smith}, {Song}, {Tolls}, {Unwin}, {Venumadhav}, {Viero},
  {Werner} \& {Zemcov}}]{Dore2014}
{Dor{\'e}}, O., {Bock}, J., {Ashby}, M., {Capak}, P., {Cooray}, A., {de
  Putter}, R., {Eifler}, T., {Flagey}, N., {Gong}, Y., {Habib}, S., {Heitmann},
  K., {Hirata}, C., {Jeong}, W.-S., {Katti}, R., {Korngut}, P., {Krause}, E.,
  {Lee}, D.-H., {Masters}, D., {Mauskopf}, P., {Melnick}, G., {Mennesson}, B.,
  {Nguyen}, H., {{\"O}berg}, K., {Pullen}, A., {Raccanelli}, A., {Smith}, R.,
  {Song}, Y.-S., {Tolls}, V., {Unwin}, S., {Venumadhav}, T., {Viero}, M.,
  {Werner}, M. \& {Zemcov}, M. [2014]  \emph{ArXiv e-prints 1412.4872} .

\bibitem[{{Dressel}(2016)}]{Dressel2016}
{Dressel}, L. [2016]  \emph{{Wide Field Camera 3 Instrument Handbook v. 8.0}}
  (Space Telescope Science Institute).

\bibitem[{{Fixsen} \emph{et~al.}(2000){Fixsen}, {Offenberg}, {Hanisch},
  {Mather}, {Nieto-Santisteban}, {Sengupta} \& {Stockman}}]{Fixsen2000}
{Fixsen}, D.~J., {Offenberg}, J.~D., {Hanisch}, R.~J., {Mather}, J.~C.,
  {Nieto-Santisteban}, M.~A., {Sengupta}, R. \& {Stockman}, H.~S. [2000]
  \emph{\pasp} \textbf{112},  1350, \doi{10.1086/316626}.

\bibitem[{{Fowler} \& {Gatley}(1990)}]{Fowler1990}
{Fowler}, A.~M. \& {Gatley}, I. [1990]  \emph{\apjl} \textbf{353},  L33,
  \doi{10.1086/185701}.

\bibitem[{{Garnett} \& {Forrest}(1993)}]{Garnett1993}
{Garnett}, J.~D. \& {Forrest}, W.~J. [1993]  \enquote{{Multiply sampled
  read-limited and background-limited noise performance},}  \emph{Infrared
  Detectors and Instrumentation}, ed. {Fowler}, A.~M. (\procspie), p. 395.

\bibitem[{{Kelsall} \emph{et~al.}(1998){Kelsall}, {Weiland}, {Franz}, {Reach},
  {Arendt}, {Dwek}, {Freudenreich}, {Hauser}, {Moseley}, {Odegard},
  {Silverberg} \& {Wright}}]{Kelsall1998}
{Kelsall}, T., {Weiland}, J.~L., {Franz}, B.~A., {Reach}, W.~T., {Arendt},
  R.~G., {Dwek}, E., {Freudenreich}, H.~T., {Hauser}, M.~G., {Moseley}, S.~H.,
  {Odegard}, N.~P., {Silverberg}, R.~F. \& {Wright}, E.~L. [1998]  \emph{\apj}
  \textbf{508},  44, \doi{10.1086/306380}.

\bibitem[{{Loose} \emph{et~al.}(2006){Loose}, {Beletic}, {Blackwell}, {Hall} \&
  {Jacobsen}}]{Loose2006}
{Loose}, M., {Beletic}, J.~W., {Blackwell}, J., {Hall}, D. \& {Jacobsen}, S.
  [2006]  \enquote{{SIDECAR ASIC - Control Electronics on a Chip},}
  \emph{Astrophysics and Space Science Library}, eds. {Beletic}, J.~E.,
  {Beletic}, J.~W. \& {Amico}, P., p. 699, \doi{10.1007/978-1-4020-4330-7_85}.

\bibitem[{{Moseley} \emph{et~al.}(2010){Moseley}, {Arendt}, {Fixsen},
  {Lindler}, {Loose} \& {Rauscher}}]{Moseley2010}
{Moseley}, S.~H., {Arendt}, R.~G., {Fixsen}, D.~J., {Lindler}, D., {Loose}, M.
  \& {Rauscher}, B.~J. [2010]  \enquote{{Reducing the read noise of H2RG
  detector arrays: eliminating correlated noise with efficient use of reference
  signals},}  \emph{High Energy, Optical, and Infrared Detectors for Astronomy
  IV}, p. 77421B, \doi{10.1117/12.866773}.

\bibitem[{{Offenberg} \emph{et~al.}(2005){Offenberg}, {Fixsen} \&
  {Mather}}]{Offenberg2005}
{Offenberg}, J.~D., {Fixsen}, D.~J. \& {Mather}, J.~C. [2005]  \emph{\pasp}
  \textbf{117},  94, \doi{10.1086/427566}.

\bibitem[{{Press} \emph{et~al.}(1992){Press}, {Teukolsky}, {Vetterling} \&
  {Flannery}}]{Press1992}
{Press}, W.~H., {Teukolsky}, S.~A., {Vetterling}, W.~T. \& {Flannery}, B.~P.
  [1992]  \emph{{Numerical recipes in C. The art of scientific computing, 2nd
  ed.}} (Cambridge University Press).

\bibitem[{{Rauscher}(2015)}]{Rauscher2015}
{Rauscher}, B.~J. [2015]  \emph{\pasp} \textbf{127},  1144,
  \doi{10.1086/684082}.

\bibitem[{{Rauscher} \emph{et~al.}(2007){Rauscher}, {Alexander}, {Brambora},
  {Derro}, {Engler}, {Fox}, {Garrison}, {Henegar}, {Hill}, {Johnson},
  {Lindler}, {Manthripragada}, {Marshall}, {Mott}, {Parr}, {Roher},
  {Shakoorzadeh}, {Smith}, {Waczynski}, {Wen}, {Wilson}, {Xia-Serafino},
  {Cabelli}, {Cheng}, {Garnett}, {Loose}, {Zandian}, {Zino}, {Ellis}, {Howe},
  {Jurado}, {Lee}, {Nieznanski}, {Wallis}, {York}, {Regan}, {Bagnasco},
  {B{\"o}ker}, {De Marchi}, {Ferruit}, {Jakobsen} \& {Strada}}]{Rauscher2007}
{Rauscher}, B.~J., {Alexander}, D., {Brambora}, C.~K., {Derro}, R., {Engler},
  C., {Fox}, O., {Garrison}, M.~B., {Henegar}, G., {Hill}, R.~J., {Johnson},
  T., {Lindler}, D.~J., {Manthripragada}, S.~S., {Marshall}, C., {Mott}, B.,
  {Parr}, T.~M., {Roher}, W.~D., {Shakoorzadeh}, K.~B., {Smith}, M.,
  {Waczynski}, A., {Wen}, Y., {Wilson}, D., {Xia-Serafino}, W., {Cabelli}, C.,
  {Cheng}, E., {Garnett}, J., {Loose}, M., {Zandian}, M., {Zino}, J., {Ellis},
  T., {Howe}, B., {Jurado}, M., {Lee}, G., {Nieznanski}, J., {Wallis}, P.,
  {York}, J., {Regan}, M.~W., {Bagnasco}, G., {B{\"o}ker}, T., {De Marchi}, G.,
  {Ferruit}, P., {Jakobsen}, P. \& {Strada}, P. [2007]  \enquote{{Detector
  arrays for the James Webb Space Telescope near-infrared spectrograph},}
  \emph{Focal Plane Arrays for Space Telescopes III}, p. 66900M,
  \doi{10.1117/12.731759}.

\bibitem[{{Robberto}(2007)}]{Robberto2007}
{Robberto}, M. [2007]  \enquote{{Analysis of the sampling schemes for
  WFC3-IR},} Tech. rep.

\bibitem[{{Smith} \& {Hale}(2012)}]{Smith2012}
{Smith}, R.~M. \& {Hale}, D. [2012]  \enquote{{Read noise for a 2.5{$\mu$}m
  cutoff Teledyne H2RG at 1-1000Hz frame rates},}  \emph{High Energy, Optical,
  and Infrared Detectors for Astronomy V}, p. 84530Y, \doi{10.1117/12.927148}.

\bibitem[{{Spangelo} \emph{et~al.}(2015){Spangelo}, {Katti}, {Unwin} \&
  {Bock}}]{Spangelo2015}
{Spangelo}, S.~C., {Katti}, R.~M., {Unwin}, S.~C. \& {Bock}, J.~J. [2015]
  \emph{Journal of Astronomical Telescopes, Instruments, and Systems}
  \textbf{1}, 037001, \doi{10.1117/1.JATIS.1.3.037001}.

\bibitem[{{Wright} \emph{et~al.}(2010){Wright}, {Eisenhardt}, {Mainzer},
  {Ressler}, {Cutri}, {Jarrett}, {Kirkpatrick}, {Padgett}, {McMillan},
  {Skrutskie}, {Stanford}, {Cohen}, {Walker}, {Mather}, {Leisawitz}, {Gautier},
  {McLean}, {Benford}, {Lonsdale}, {Blain}, {Mendez}, {Irace}, {Duval}, {Liu},
  {Royer}, {Heinrichsen}, {Howard}, {Shannon}, {Kendall}, {Walsh}, {Larsen},
  {Cardon}, {Schick}, {Schwalm}, {Abid}, {Fabinsky}, {Naes} \&
  {Tsai}}]{Wright2010}
{Wright}, E.~L., {Eisenhardt}, P.~R.~M., {Mainzer}, A.~K., {Ressler}, M.~E.,
  {Cutri}, R.~M., {Jarrett}, T., {Kirkpatrick}, J.~D., {Padgett}, D.,
  {McMillan}, R.~S., {Skrutskie}, M., {Stanford}, S.~A., {Cohen}, M., {Walker},
  R.~G., {Mather}, J.~C., {Leisawitz}, D., {Gautier}, T.~N., III, {McLean}, I.,
  {Benford}, D., {Lonsdale}, C.~J., {Blain}, A., {Mendez}, B., {Irace}, W.~R.,
  {Duval}, V., {Liu}, F., {Royer}, D., {Heinrichsen}, I., {Howard}, J.,
  {Shannon}, M., {Kendall}, M., {Walsh}, A.~L., {Larsen}, M., {Cardon}, J.~G.,
  {Schick}, S., {Schwalm}, M., {Abid}, M., {Fabinsky}, B., {Naes}, L. \&
  {Tsai}, C.-W. [2010]  \emph{\aj} \textbf{140}, 1868-1881,
  \doi{10.1088/0004-6256/140/6/1868}.

\bibitem[{{Xilinx}(2012)}]{Virtex5}
{Xilinx} [2012]  \emph{{Virtex-5 FPGA User Guide}} (Xilinx, Inc.).

\bibitem[{{Zemcov} \emph{et~al.}(2014){Zemcov}, {Smidt}, {Arai}, {Bock},
  {Cooray}, {Gong}, {Kim}, {Korngut}, {Lam}, {Lee}, {Matsumoto}, {Matsuura},
  {Nam}, {Roudier}, {Tsumura} \& {Wada}}]{Zemcov2014}
{Zemcov}, M., {Smidt}, J., {Arai}, T., {Bock}, J., {Cooray}, A., {Gong}, Y.,
  {Kim}, M.~G., {Korngut}, K., {Lam}, A., {Lee}, D.~H., {Matsumoto}, T.,
  {Matsuura}, S., {Nam}, U.~W., {Roudier}, G., {Tsumura}, K. \& {Wada}, T.
  [2014]  \emph{Science} \textbf{346},  6210.

\end{thebibliography}

\end{document}